\documentclass[
 reprint,
superscriptaddress,
 amsmath,amssymb,
 aps,nofootinbib,
 prd
]{revtex4-2}

\usepackage{graphicx}
\usepackage[usenames,dvipsnames]{xcolor}
\definecolor{bluecite}{HTML}{0875b7}
\usepackage[colorlinks=true,urlcolor=Emerald,linkcolor=bluecite,citecolor=bluecite]{hyperref}
\usepackage{acronym}

\newcommand{\order}[1]{{\cal O}(#1)}

\newcommand{\fderiv}[2]{\frac{\delta #1}{\delta #2}}
\newcommand{\dd}{\text{d}}

\newcommand{\covd}{\ensuremath{\nabla}}
\newcommand{\bcovd}{\ensuremath{\bar{\covd}}}
\newcommand{\bDelta}{\ensuremath{\bar{\Delta}}}
\newcommand{\bR}{\ensuremath{\bar{R}}}
\newcommand{\bC}{\ensuremath{\bar{C}}}
\newcommand{\bK}{\ensuremath{\bar{K}}}
\newcommand{\bMcal}{\ensuremath{\bar{\mathcal M}}}
\newcommand{\bPi}{\ensuremath{\bar{\Pi}}}
\newcommand{\bg}{\ensuremath{\bar{g}}}
\newcommand{\dimlessgamma}{\ensuremath{\tilde\gamma}}

\newcommand{\eg}{{\textit{e.g.}}}
\newcommand{\ie}{{\textit{i.e.}}}

\newcommand{\eulerterm}{\ensuremath{\mathfrak E}}

\newacro{UV}[UV]{ultraviolet}
\newacro{IR}[IR]{infrared}
\newacro{QFT}[QFT]{quantum field theory}
\newacro{EFT}[EFT]{effective field theory}
\newacro{GR}[GR]{General Relativity}
\newacro{FRG}[FRG]{Functional Renormalisation Group}
\newacro{RG}[RG]{renormalisation group}
\newacro{EAA}[EAA]{effective average action}
\newacro{PMS}[PMS]{principle of minimal sensitivity}
\newacro{MES}[MES]{minimal essential scheme}

\begin{document}

\title{Robustness of the derivative expansion in Asymptotic Safety}

\author{Alessio Baldazzi\,\href{https://orcid.org/0000-0001-7075-6047}{\protect \includegraphics[scale=.07]{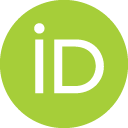}}\,}
\email{alessio.baldazzi@unitn.it}
\affiliation{Department of Physics, University of Trento, Via Sommarive 14, 38123, Trento, Italy}

\author{Kevin Falls\,\href{https://orcid.org/0000-0003-1265-0492}{\protect \includegraphics[scale=.07]{ORCIDiD_icon128x128.png}}\,}
\email{kfalls@fing.edu.uy}
\affiliation{Instituto de F\'isica, Facultad de Ingenier\'ia, Universidad de la Rep\'ublica, J.H.y Reissig 565, 11300 Montevideo, Uruguay}

\author{Yannick Kluth\,\href{https://orcid.org/0000-0002-8126-7668}{\protect \includegraphics[scale=.07]{ORCIDiD_icon128x128.png}}\,}
\email{yannick.kluth@manchester.ac.uk}
\affiliation{
 Department of Physics and Astronomy, University of Manchester, Manchester M13 9PL, United Kingdom
}
\affiliation{Perimeter Institute for Theoretical Physics, Waterloo, Ontario N2L 2Y5, Canada}%

\author{Benjamin Knorr\,\href{https://orcid.org/0000-0001-6700-6501}{\protect \includegraphics[scale=.07]{ORCIDiD_icon128x128.png}}\,}
\email[]{benjamin.knorr@su.se}
\affiliation{Nordita, Stockholm University and KTH Royal Institute of Technology, Hannes Alfv\'ens v\"ag 12, SE-106 91 Stockholm, Sweden}

\date{\today}

\begin{abstract}
We analyse the renormalisation group flow of quantum gravity at sixth order in the derivative expansion within the background field approximation.
Non-linear field redefinitions are used to ensure that only essential couplings flow. Working within the universality class of General Relativity, with a vanishing cosmological constant,  redundant couplings are fixed to their values at the Gaussian fixed point.
This reduces the theory space to two dynamical essential couplings given by Newton's and the Goroff-Sagnotti coupling. Furthermore, it implements the condition that no extra degrees of freedom are present beyond those of General Relativity, in contrast to higher derivative theories and derivative expansions in a conventional renormalisation scheme. We find a unique ultraviolet fixed point with a single relevant direction and analyse the phase diagram of the theory. Our results suggest resilience of the gravitational Reuter fixed point under the inclusion of higher order curvature invariants and show several signs of near-perturbativity.
The regulator dependence of our results is investigated in detail and shows that qualitative and quantitative features are robust to a large extent.
\end{abstract}
\maketitle

\tableofcontents

\section{Introduction}\label{sec:intro}

Finding a quantum theory of gravity is one of the most fascinating open problems in theoretical physics. Such a theory is supposed to shine light on fundamental questions about the origin of the universe and the micro-structure of black holes. 
However, pure Einstein gravity is well-known to be perturbatively non-renormalisable due to Newton's coupling having negative mass dimension. As a result, perturbative renormalisability fails from two loops onwards as signalled by the presence of the Goroff-Sagnotti counterterm \cite{Goroff:1985sz,Goroff:1985th,VANDEVEN}. This indicates that pertubative Einstein quantum gravity has to include an infinite number of parameters, rendering the theory non-predictive. Below the Planck scale, predictivity can be regained by treating gravity as an \ac{EFT} \cite{Donoghue:1994dn}. In this regime, higher order couplings are suppressed by powers of the Planck mass and can be neglected. The remaining finite number of couplings can be fixed by observations, at least in principle \cite{Donoghue:1994dn}. However, this approach breaks down at the Planck scale and requires an \ac{UV} completion.

The Asymptotic Safety scenario has emerged as one of the key contenders for a viable \ac{UV} completion of quantum gravity \cite{Weinberg:1976xy,Weinberg:1980gg,Parisi:1977uz}. It is based entirely on standard \ac{QFT}, and suggests that quantum gravity becomes scale-invariant at short distances. This can be realised by an interacting fixed point of the \ac{RG} flow. Evidence for the existence of such a fixed point has accumulated, both in pure gravity, where it is known as the Reuter fixed point \cite{Reuter_1998}, and when coupled to matter, see \eg{} the collection of book chapters \cite{Knorr:2022dsx, Eichhorn:2022gku, Morris:2022btf, Martini:2022sll, Wetterich:2022ncl, Platania:2023srt, Saueressig:2023irs, Pawlowski:2023gym} for an up-to-date overview of the field. For some recent work in the field see \eg{} 
\cite{Martini:2023qkp, Bonanno:2023fij, Bonanno:2023ghc, DAngelo:2023wje, Pawlowski:2023dda, Eichhorn:2023jyr, Bonanno:2023rzk, deBrito:2023myf, Saueressig:2023tfy, Kawai:2023rgy, Knorr:2023usb, deBrito:2023kow,deBrito:2023ydd}.

Since Asymptotic Safety is a non-perturbative phenomenon, small coupling expansions cannot be relied on and a mathematical proof for the existence of a fixed point seems out of reach for the moment. Nevertheless, there are strong indications that Asymptotic Safety is realised in a near-perturbative regime: it is as non-perturbative as it needs to be, but as perturbative as it can be \cite{Falls:2013bv,Falls:2014tra, Eichhorn:2018akn, Eichhorn:2018ydy}. This gives some evidence that central aspects of the theory are already well-captured by current approximations.

In this work, we provide more evidence for this observation. Specifically, we investigate Asymptotic Safety in a derivative expansion up to sixth order on a general background, for the first time considering a complete basis of operators at this order. This goes significantly beyond previous work, and breaks a barrier in terms of technical feasibility. Previous computations were either limited to fourth order in derivatives \cite{Falls:2020qhj,Knorr:2021slg,Sen:2021ffc}, or to selecting very specific backgrounds. Consequently, this lead to either an identification of operators with an equal number of derivatives, or to a straight neglect of certain operators \cite{Codello:2008vh,Benedetti:2009gn,Falls:2014tra, Gies:2016con, Falls:2017lst, Kluth:2020bdv, Kluth:2022vnq}.

In our endeavour, we rely on the freedom to perform field redefinitions of the metric to fix all inessential couplings at all scales. This follows recent technical progress in the field \cite{Pawlowski:2005xe, Baldazzi:2021ydj, Baldazzi:2021orb, Knorr:2022ilz}. Apart from technical simplifications, the resulting clear distinction between essential and inessential couplings is beneficial when counting the number of free parameters in the theory. Moreover, it imposes that no extra degrees of freedom are present apart from those included in \ac{GR}. This is contrary to higher derivative theories such as quadratic gravity or derivative expansions in the standard renormalisation scheme, in which higher order curvatures lead to the presence of (fiducial or actual) ghosts \cite{Stelle:1977ry, Platania:2020knd, Platania:2022gtt}.

This work is structured as follows.
In \autoref{sec:EFRforgra}, preliminaries including the usage of field redefinitions at sixth order in derivatives for pure gravity are explained. The results of our computation are presented in \autoref{sec:results}. We discuss properties of the \ac{UV} fixed point (\autoref{sec:uvfp}), the phase diagram (\autoref{sec:RGglowandIR}) and regulator dependence (\autoref{sec:regdep}-\autoref{sec:regdepbeyond}). In \autoref{sec:discussion} we summarise the main results and give future perspectives. In appendix \ref{app:C3standard}, we refer to an earlier computation including the Goroff-Sagnotti term in the standard scheme, while appendix \ref{app:GB} contains some remarks about the topological Gauss-Bonnet term.

\section{Essential renormalisation group for gravity}\label{sec:EFRforgra}

In this section, we discuss some basics of implementing field redefinitions in non-perturbative \ac{RG} flows as well as our setup and technical choices.
\\

\subsection{Generalised flow equations}\label{sec:genFRGflow}

In order to study the non-perturbative renormalisability of quantum gravity, we employ the \ac{FRG} \cite{Wetterich:1992yh, Morris:1993qb, Ellwanger:1993mw}.
It realises Wilson's formulation of the \ac{RG} whereby fluctuations are integrated out with respect to a fiducial momentum scale $k$. 
For every scale $k$, this gives rise to an \ac{EAA} $\Gamma_k[\Phi]$ that describes physics at the scale $k$, where modes with momentum $p^2>k^2$ have been integrated out.
In the generalised formulation \cite{Pawlowski:2005xe,Wegner1974,Baldazzi:2021ydj}, we allow for the freedom to change the description of physics at different scales $k$ by means of local field redefinitions.
Thus, the dependence on $k$ originates from the Wilsonian \ac{IR} cutoff and from the $k$-dependent choice of field variables. Explicitly, the \ac{EAA} can be defined by the integro-differential equation
\begin{widetext}
\begin{equation} 
    e^{-\Gamma_k[\Phi]} = \int [D\chi] e^{-S[\chi] + ( \hat{\Phi}_k[\chi] - \Phi)^A \frac{\delta}{\delta \Phi^A} \Gamma[\Phi]    - \frac{1}{2} ( \hat{\Phi}_k[\chi] - \Phi)^A [\mathcal{R}]_{AB} ( \hat{\Phi}_k[\chi] - \Phi)^B  } \, .
    \label{eqn:EAA}
\end{equation}
\end{widetext}
We use DeWitt notation with a repeated index implying both a sum over field indices and an integration over spacetime.
The argument of the \ac{EAA} is the expectation value of the microscopic field $\Phi = \langle \hat{\Phi}_k \rangle_{k,\Phi}$. Exploiting the freedom to choose the microscopic field variables $\hat{\Phi}_k[\chi]$ as a function of $k$ allows for general non-linear local field redefinitions. These enter the flow equation for the \ac{EAA} via the \ac{RG} kernel $\Psi_k$, which is defined as the expectation value of the $k$-derivative of the microscopic field,
\begin{equation}
\Psi_k = \langle \partial_t \hat{\Phi}_k \rangle_{k,\Phi}\,,
\end{equation}
where $t = \log(k/k_0)$ is the \ac{RG} time and $k_0$ an arbitrary reference scale.

The \ac{EAA} obeys the exact functional flow equation \cite{Pawlowski:2005xe}
\begin{widetext}
\begin{equation} 
    \left( \partial_t + \Psi^A \fderiv{}{\Phi^A} \right) \Gamma_k = \frac{1}{2} (-1)^{[A]} \left[\left(\Gamma_k^{(2)} + \mathcal{R}_k \right)^{-1} \right]^{A B} \left( \partial_t \delta^C_{\ B} + 2 \fderiv{\Psi^C}{\Phi^B} \right) \left[ \mathcal{R}_k \right]_{C A} \, ,
    \label{eqn:essRG}
\end{equation}
\end{widetext}
where $[A]$ takes values $1$ or $-1$ depending on whether the field corresponding to the index $A$ is commuting or anti-commuting. 
The flow equation depends on the \ac{IR} cutoff $\mathcal{R}_k$, which implements a gradual integration of fluctuations in the path integral, and the \ac{RG} kernel $\Psi_k$, which parameterises field redefinitions.

In the case of pure gravity, the fields $\Phi^A$ include both the metric tensor $g_{\mu\nu}$ and the Faddeev-Popov ghosts $c_\mu$ and $\bar{c}_\mu$. Furthermore, the \ac{EAA} depends on a background metric $\bg_{\mu\nu}$ introduced both to fix the gauge and to employ a covariant regularisation. The total metric is split into the background plus a fluctuation field $h_{\mu \nu}$,
\begin{equation}
g_{\mu\nu} = \bg_{\mu\nu} + h_{\mu\nu} \, .
\end{equation}

In this paper, we use the background field approximation, where $\Gamma_k[\Phi]$ is approximated by a diffeomorphism invariant functional $\bar{\Gamma}_k[g]$ in addition to gauge fixing and ghost terms,
\begin{equation}
\Gamma_k = \bar{\Gamma}_k[g] +  S_\textrm{gf}[g; \bg ]  + S_\textrm{gh}[g,c,\bar{c}; \bg ] \,.
\end{equation}
To close the background field approximation, we identify the full metric $g_{\mu \nu}$ with the background metric $\bg_{\mu \nu}$ by setting $h_{\mu\nu}=0$ in \eqref{eqn:essRG}.
One then obtains the $\beta$-functions for the couplings in $\bar{\Gamma}_k[g]$ which multiply diffeomorphism invariant operators.
Importantly, as we indicate below, the gauge fixing and ghost terms in the action depend on $k$ only through Newton's coupling $G_N= G_N(k)$. This reflects that we study the universality class with the same degrees of freedom as \ac{GR}.

\subsection{Minimal Essential Scheme at sixth order}\label{sec:scheme_cubic}

In this paper, we implement the \ac{MES} for gravity \cite{Baldazzi:2021ydj,Baldazzi:2021orb}, which consists of putting to zero all terms that are proportional to the vacuum Einstein equations, $R_{\mu\nu}=0$, apart from the Einstein-Hilbert term.
This can be achieved within a consistent approximation where the flow equation is expanded in derivatives with an appropriate choice of the \ac{RG} kernel $\Psi_k$. In particular, if we include terms with up to $n$ derivatives of the metric in the action, we include terms with $(n-2)$ derivatives in the \ac{RG} kernel. Previously, this scheme has been carried out to order $n=4$ for both pure gravity \cite{Baldazzi:2021orb} and gravity coupled to a shift-symmetric scalar field \cite{Knorr:2022ilz}. We extend this to the sixth order ($n = 6$), whence the \ac{EAA} reads
\begin{equation}
    \begin{aligned}
        \bar{\Gamma}_k = \int \dd^4x \sqrt{g} \bigg[\, \frac{\rho}{8 \pi} \, -\,  &\frac{1}{16\pi G_N} R + \sigma_\eulerterm \eulerterm \\
        & + G_{C^3} C^{\rho \sigma}_{\phantom{\rho\sigma} \mu \nu} C^{\mu \nu}_{\phantom{\mu\nu} \alpha \beta} C^{\alpha \beta}_{\phantom{\alpha\beta} \rho \sigma} \bigg] \, .
        \label{eq:barGammaform}
    \end{aligned}
\end{equation}
Here, $\rho =  \Lambda/G_N$ gives the vacuum energy of the field, $\eulerterm = R^2 - 4 R_{\mu \nu} R^{\mu \nu} + R_{\rho \sigma \mu \nu} R^{\rho \sigma \mu \nu}$ is the topological Gauss-Bonnet term, and $C_{\rho \sigma \mu \nu}$ is the Weyl tensor. In addition to Newton's coupling $G_N$ and the topological Gauss-Bonnet coupling $\sigma_\eulerterm$, the Goroff-Sagnotti coupling $G_{C^3}$ \cite{Goroff:1985sz,Goroff:1985th} enters as the second dynamical essential coupling. All other couplings to operators with up to six derivatives are set to zero.

An important additional element of the \ac{MES} for gravity is the treatment of the cosmological constant \cite{Baldazzi:2021orb}. The scheme specifies that $\rho$ is fixed to be proportional to the fourth power of the cutoff scale $k$. The constant of proportionality is fixed at the Gaussian fixed point and depends on the regulator. This imposes that the physical cosmological constant vanishes, \ie{} when $k$ goes to zero. Moreover, we note that this condition also dictates that the dimensionless product $G_N \Lambda$ is given by $G_N \Lambda = \tilde{\rho} \, k^4 G_N^2$, where the proportionality constant $\tilde{\rho}$ depends on the regularisation. The vacuum energy has also been fixed in an alternative fashion in \cite{Kawai:2023rgy} which involves rescaling the metric and modifying the regulator.
We also note that using certain gauges and parameterisation of the metric the cosmological constant can be removed from the trace part flow equation \cite{Ohta:2015efa,Ohta:2015fcu,Falls:2015qga}. Furthermore in unimodular gravity the Newton's constant is essential and the cosmological constant appears to play no role \cite{Benedetti:2015zsw}.
For a more general discussion on the redundant nature of the cosmological constant see \cite{Hamber:2013rb}.

The \ac{MES} for gravity fixes the inessential couplings to zero as they are identified at the Gaussian fixed point, defined by taking each coupling $G_N$, $\rho$ and $G_{C^3}$ to zero. Away from the Gaussian fixed point we continue to fix the inessential couplings which are those conjugate to redundant operators that are given by
\begin{equation} 
\begin{split}
\mathcal{T}[\Xi_g] &:= \, \Xi_g^A  \frac{\delta }{\delta g^A}  \bar{\Gamma}_k \\
&\qquad - \left[\left(\bar{\Gamma}_k^{(2)} + \mathcal{R}_k \right)^{-1} \right]^{A B} \fderiv{\Xi^C}{g^B} \left[ \mathcal{R}_k \right]_{C A}   \,,
\end{split}
\label{eq:reduope}
\end{equation}
where $\Xi_g $ are symmetric covariant tensors composed of the metric and its derivatives.\footnote{We do not allow for redefinitions of the ghosts since we do not compute the flow of their couplings in our approximation.}
Maintaining the form of the action \eqref{eq:barGammaform} along the flow continues to fix the inessential couplings provided that the redundant operators form a complete basis with the essential operators \cite{Baldazzi:2021ydj,Baldazzi:2021orb}.  
At sixth order in the derivative expansion, the redundant operator basis is determined by the following linearly independent set of $\Xi^g_{\mu\nu}$ inserted in \eqref{eq:reduope}: 
\begin{equation}
\begin{split}
&\Big\{ g_{\mu\nu}, \, R g_{\mu\nu},\, S_{\mu\nu},\,
    R^2 \, g_{\mu \nu},\,
     S_{\rho \sigma} S^{\rho \sigma} g_{\mu \nu} 
     ,\\
    & \quad C_{\rho \sigma \alpha \beta} C^{\rho \sigma \alpha \beta} g_{\mu \nu} 
    ,\, S_{\mu \rho} S^\rho_{\phantom{\rho} \nu} 
    ,\,  C_{\mu \rho \nu \sigma} S^{\rho \sigma}, \\
    & \quad g_{\mu \nu} \, \Delta R ,\, \Delta S_{\mu \nu}, \,
    R \, S_{\mu \nu} ,\, \covd_\mu \covd_\nu R
\Big\}
    \,.
    \label{eq:Gredu}
\end{split}
\end{equation}
Here, $\Delta=-\nabla^2$ and $S_{\mu\nu} = R_{\mu\nu} - \frac{1}{4} g_{\mu\nu} R$ is the trace-free Ricci tensor. A general ansatz for the \ac{RG} Kernel is set by the choice in \eqref{eq:Gredu} and is given by
\begin{equation}
    \begin{aligned}
        \Psi^{g}_{\mu\nu} =& \, \gamma_g g_{\mu \nu} \\
        & + \gamma_R \, R \, g_{\mu \nu} + \gamma_{S} \, S_{\mu \nu} \\
        & + \gamma_{R^2} \, R^2 \, g_{\mu \nu} \\
        & + \gamma_{S^2} \left( S_{\rho \sigma} S^{\rho \sigma} g_{\mu \nu} + (R-4\Lambda) S_{\mu\nu} \right) \\
        & + \gamma_{C^2} \left( C_{\rho \sigma \alpha \beta} C^{\rho \sigma \alpha \beta} g_{\mu \nu} - 8 \Lambda S_{\mu\nu} \right) \\
        & + \gamma_{SSTL} \left(S_{\mu \rho} S^\rho_{\phantom{\rho} \nu} - \frac{1}{4}\,g_{\mu \nu} S^{\rho \sigma} S_{\rho \sigma} \right) \\
        & + \gamma_{CS} \, C_{\mu \rho \nu \sigma} S^{\rho \sigma} \\
        & + \gamma_{\Delta R} \, g_{\mu \nu} \, \Delta R + \gamma_{\Delta S} \, \Delta S_{\mu \nu} \\
        & + \gamma_{R S} \, R \, S_{\mu \nu} + \gamma_{DDR} \, \covd_\mu \covd_\nu R \\
        & + \mathcal{O} \left(\partial^6\right) \, .
    \end{aligned}   
    \label{eqn:RGkernel}
\end{equation}
The explicit dependence on the cosmological constant $\Lambda = G_N \rho$ is simply introduced for convenience -- it brings the classical contribution of the \ac{RG} kernel to the flow equation (the contribution on the left-hand side of \eqref{eqn:essRG}) into an almost diagonal form for our chosen basis of curvature invariants,
\begin{equation}
    \begin{split}
        \Psi_{\mu \nu}^g \fderiv{\bar{\Gamma}_k}{g_{\mu \nu}}
        =& \, \int \dd^4 x \sqrt{g} \, \frac{1}{16 \pi G_N} \times \\
        & \Big[ 4 \Lambda \gamma_g + \left(4 \Lambda \gamma_R - \gamma_g  \right) R \\
        & + \left(- \gamma_R + 4 \Lambda \gamma_{R^2} - \frac{2}{3} \Lambda \gamma_{C^2} \right) R^2 \\
        & + \gamma_S S_{\mu \nu} S^{\mu \nu} + 4 \Lambda \gamma_{C^2} \eulerterm - \gamma_{R^2} R^3 \\
        & + \gamma_{RS} R S_{\mu \nu} S^{\mu \nu} - \gamma_{C^2} R C_{\rho \sigma \mu \nu} C^{\rho \sigma \mu \nu} \\
        & + \gamma_{SSTL} S^\mu_{\phantom{\mu} \nu} S^\nu_{\phantom{\nu} \rho} S^\rho_{\phantom{\rho} \mu} + \gamma_{CS} S_{\mu \nu} S_{\rho \sigma} C^{\mu \rho \nu \sigma} \\
        & - 16 \pi G_N \gamma_g G_{C^3} C^{\rho \sigma}_{\phantom{\rho\sigma} \mu \nu} C^{\mu \nu}_{\phantom{\mu\nu} \alpha \beta} C^{\alpha \beta}_{\phantom{\alpha\beta} \rho \sigma} \\
        & - \gamma_{\Delta R} R \Delta R + \gamma_{\Delta S} S_{\mu \nu} \Delta S^{\mu \nu} \Big] \, .
    \end{split}
    \label{eqn:rgclass}
\end{equation}
From this expression, we can see explicitly that we can choose the $\gamma$-functions in such a way that we do not generate inessential couplings along the flow.
Note that the terms proportional to $\gamma_{DDR}$ and $\gamma_{S^2}$ of the \ac{RG} kernel \eqref{eqn:RGkernel} do not contribute to \eqref{eqn:rgclass}. This fact is explained in the following subsection.

\subsection{Symmetries of the RG kernel}\label{sec:RGsym}

The \ac{RG} kernel as defined in \eqref{eqn:RGkernel} includes terms that represent symmetries of the underlying action. Due to this fact, they do not lead to contributions to the left-hand side of the essential \ac{RG} equation \eqref{eqn:essRG}, as seen in \eqref{eqn:rgclass}. One example for such terms is $\gamma_{DDR} \covd_\mu \covd_\nu R$. Since
\begin{equation}
    \covd_\mu \xi_\nu + \covd_\nu \xi_\mu = \covd_\mu \covd_\nu R \, ,
\end{equation}
with $\xi_\mu = \covd_\mu R/2$, this term is a diffeomorphism and does not contribute to the left-hand side of \eqref{eqn:essRG}. However, it does lead to a non-vanishing contribution on the right-hand side. Moreover, its contribution to the left-hand side of \eqref{eqn:essRG} only vanishes within the background field approximation. If non-diffeomorphism invariant terms are included in the action, there is also a non-trivial contribution to the left-hand side. We conclude that the corresponding tree-level redundant operator is outside of our approximation. Therefore, we set the term $ \gamma_{DDR} \covd_\mu \covd_\nu R$ to zero in our ansatz for $\Psi_k$. The same reasoning applies to other diffeomorphism invariant contributions to the \ac{RG} kernel that would be encountered at higher orders of the derivative expansion.

Another example for a term of the \ac{RG} kernel that does not contribute to the left-hand side of \eqref{eqn:essRG} is
\begin{equation}
 \gamma_{S^2} \left( S_{\rho \sigma} S^{\rho \sigma} g_{\mu \nu} + (R-4\Lambda) S_{\mu\nu} \right) \, .
 \label{eqn:gammaSsq}
\end{equation}
In contrast to $ \gamma_{DDR} \covd_\mu \covd_\nu R$, it is not a diffeomorphism term, but rather an example of another class of symmetries \cite{DeWitt:2003pm}.
Any contribution to the \ac{RG} kernel of the form
\begin{equation}\label{eq:Psi_sym}
    \Psi^{\text{sym}}_{\mu\nu} (x) = \int \dd^4 y \, \sqrt{g} \, T_{\mu \nu, \rho \sigma} (x, y) \, \frac{\delta \Gamma}{\delta g_{\rho \sigma} (y)}
\end{equation}
is a symmetry, provided that the two-point function $T_{\mu\nu, \rho \sigma} (x, y)$ satisfies $T_{\mu\nu, \rho \sigma} (x, y) = - T_{\rho \sigma, \mu\nu} (y, x)$. Note that such contributions are not specific to gravity, but also arise in other theories, \eg{}, in scalar field theories. In the case of \eqref{eqn:gammaSsq}, we have
\begin{equation}
    T_{\mu \nu, \rho \sigma} (x, y) = \delta (x - y) \left[g_{\mu \nu} (x) S_{\rho \sigma} (y) - S_{\mu \nu} (x) g_{\rho \sigma} (y)\right] \, ,
\end{equation}
up to proportionality factors. Similarly to diffeomorphisms, the contribution of \eqref{eqn:gammaSsq} vanishes on the left-hand side of \eqref{eqn:essRG} within the background field approximation, while it leads to a non-vanishing contribution on the right-hand side.
However, such contributions on the right-hand side are proportional to the equations of motion of the \ac{EAA} for any term of the form \eqref{eq:Psi_sym}. This can be shown in general by noting that the contribution is proportional to the field derivative of the \ac{RG} kernel. To see this, we write the general form of \eqref{eq:Psi_sym} in DeWitt notation as
\begin{equation}
    \Psi^{\text{sym}\,C} =  T^{CD}\Gamma^{(1)}_D \,, \,\,\,\,  T^{CD}=-T^{CD} \, .
\end{equation}
Using that $T^{CD}$ is an anti-symmetric two-point function while the regulator and propagator are symmetric, it can be shown that 
\begin{equation}
    \mathcal{G}^{A B}   \fderiv{\Psi^{\text{sym}\,C}}{\Phi^B} \left[ \mathcal{R}_k \right]_{C A} =  \mathcal{G}^{A B}   \fderiv{T^{CD}  }{\Phi^B} \left[ \mathcal{R}_k \right]_{C A} \Gamma^{(1)}_D \,.
\end{equation}
In addition to this observation, we note that in the absence of a regulator, redundant operators of the effective action only have the tree-level term, and hence the redundant operator associated with \eqref{eq:Psi_sym} vanishes, but we expect the expectation values of local operators to be non-local. As a result, non-local terms in $\Psi$ can arise from local field redefinitions.

\subsection{Gauge Fixing and Regularisation}\label{sec:gaugefixing}

For practical computations utilising the \ac{FRG} in gravity, we require to fix the gauge and implement a regulator. In the \ac{MES}, we can find convenient choices for both by only considering the Einstein-Hilbert part of the action. This is due to the fact that, within the \ac{MES}, the propagator on a conformally flat spacetime coincides with that derived from \ac{GR}. 
Hence, we implement a standard background field harmonic gauge of the form
\begin{equation}
    S_\textrm{gf}[g; \bg] = \frac{1}{2} \int \dd^4x \, \sqrt{\bg}\, \chi_\mu [h, \bg] \chi^\mu [h, \bg] \, ,
    \label{eqn:essrggf}
\end{equation}
with
\begin{equation}
    \chi_\mu [h; \bg] = \frac{1}{\sqrt{16 \pi G_N}} \left[ \bcovd_\nu h^\nu_\mu - \frac{1}{2} \bcovd_\mu h \right] \,.
    \label{eqn:chi}
\end{equation}
Bars denote that respective quantities are constructed from the background metric $\bg_{\mu \nu}$.
The resulting Hessian including the Einstein-Hilbert term together with the gauge fixing is given by
\begin{equation}
    \delta^2 \left( S_\text{EH} + S_\text{gf} \right) = \frac{1}{32 \pi G_N}\int \dd^4 x \sqrt{\bg} \, h_{\mu \nu} \bDelta^{\phantom{\text{EH}}\mu \nu}_{\text{EH} \phantom{\mu\nu} \rho \sigma} h^{\rho \sigma} \, .
    \label{eqn:EHLaplace}
\end{equation}
The gauge fixing \eqref{eqn:essrggf} together with \eqref{eqn:chi} leads to a particularly simple form for the Einstein-Hilbert Laplacian $\bDelta_\text{EH}$,
\begin{equation}
    \begin{split}
        \bDelta^{\phantom{\text{EH}}\mu \nu}_{\text{EH} \phantom{\mu\nu} \rho \sigma} &=\, \left[ - \bcovd^2 - 2 \Lambda \right] \bK^{\mu\nu}_{\phantom{\mu\nu} \rho \sigma} \\
    & \quad + \left[- 2 \, \bC^{\mu\phantom{\rho}\nu}_{\phantom{\mu}\rho\phantom{\nu}\sigma} + \frac{2}{3} \, \bR \, \bPi^{\text{TL}\,\mu\nu}_{\phantom{\text{TL}\,\mu\nu}\rho\sigma} \right] \, ,
    \end{split}
    \label{eqn:EHvar}
\end{equation}
where the matrix $\bK^{\mu \nu}_{\phantom{\mu\nu}\rho \sigma}$ is given by
\begin{equation}
    \bK^{\mu\nu}_{\phantom{\mu\nu} \rho \sigma} = \bPi^{\text{TL} \, \mu \nu}_{\phantom{\text{TL}\,\mu\nu} \rho \sigma} - \bPi^{\text{Tr} \, \mu \nu}_{\phantom{\text{Tr}\,\mu\nu} \rho \sigma} \, ,
\end{equation}
and $\bPi^\text{TL}$ and $\bPi^\text{Tr}$ denote the projectors on the traceless and trace modes of the graviton, respectively,
\begin{equation}
    \begin{split}
        \bPi^{\text{TL} \, \mu \nu}_{\phantom{\text{TL}\,\mu\nu} \rho \sigma} &=\, \delta^\mu_{\ (\rho} \delta^\nu_{\ \sigma)} - \frac{1}{4} \, \bg^{\mu \nu} \bg_{\rho \sigma} \, , \\
        \bPi^{\text{Tr} \, \mu \nu}_{\phantom{\text{Tr}\,\mu\nu} \rho \sigma} &=\, \frac{1}{4} \, \bg^{\mu \nu} \bg_{\rho \sigma} \, .
    \end{split}
\end{equation}
Other choices for the gauge fixing term typically lead to non-minimal, or higher order differential operators in \eqref{eqn:EHvar}. Both are avoided in this gauge and in the \ac{MES}.

The gauge fixing \eqref{eqn:essrggf} is accompanied by the usual Faddeev-Popov ghosts in the path integral. Their contribution reads
\begin{equation}
    S_\textrm{gh} = \int \dd^4x \sqrt{\bg} \, \bar{c}_\mu \mathcal{M}^{\mu \nu} c_\nu \, ,
\end{equation}
where $c_\mu$ and $\bar{c}_\mu$ are complex Grassmann fields, and the operator $\mathcal{M}^{\mu \nu}$ is given by
\begin{equation}
    \mathcal{M}^{\mu \nu} = \fderiv{\chi^\mu [\bg, h[\xi]]}{\xi_\nu} \, .
\end{equation}
Within the background field approximation, it suffices to set $h_{\mu \nu} = 0$ in $\mathcal{M}_{\mu \nu}$. This gives rise to
\begin{equation}
    \bMcal^{\mu \nu} = \frac{1}{\sqrt{16 \pi G_N}}  \left(\bg^{\mu \nu} \bcovd^2 + \bR^{\mu \nu} \right) \, .
\end{equation}

Next, we choose a regulator for the graviton. Motivated by the form of \eqref{eqn:EHvar}, we pick
\begin{equation}
    \begin{split}
        \mathcal{R}^{g \,\mu \nu}_{k\,\phantom{\mu\nu} \rho \sigma} &= \, R_k^g \left( \bDelta \right) \bK^{\mu \nu}_{\phantom{\mu\nu} \rho \sigma} \\
        &= \, R_k^g \left( \bDelta \right) \bPi^{\text{TL}\,\mu \nu}_{\phantom{\text{TL}\,\mu\nu} \rho \sigma} - R_k^g \left( - \bcovd^2 \right) \bPi^{\text{Tr}\,\mu \nu}_{\phantom{\text{Tr}\,\mu\nu} \rho \sigma} \, .
    \end{split}
\end{equation}
The Laplacian $\bDelta$ contains a non-trivial endomorphism,
\begin{equation}
    \begin{split}
        \bDelta^{\mu \nu}_{\phantom{\mu\nu} \rho \sigma} &= \, - \bPi^{\text{TL} \, \mu \nu}_{\phantom{\text{TL}\,\mu\nu} \rho \sigma} \bcovd^2 \\
        & \quad + \beta \bigg[ - 2 \, \bC^{(\mu \phantom{\rho} \nu)}_{\phantom{(\mu} \rho \phantom{\nu)} \sigma} + \frac{2}{3} \, \bR \, \bPi^{\text{TL} \, \mu \nu}_{\phantom{\text{TL}\,\mu\nu} \rho \sigma} \bigg] \, ,
    \end{split}
    \label{eqn:endodef}
\end{equation}
that only contributes to the traceless, but not the trace mode of the graviton. The parameter $\beta$ controls the magnitude of the endomorphism. For $\beta = 0$ it vanishes, and we recover a type I cutoff. For $\beta = 1$, the Laplacian $\bDelta$ becomes equal to the Laplacian of the Einstein-Hilbert action, see \eqref{eqn:EHLaplace}. We also refer to this choice as the type II cutoff. Most of our analysis below is focused on $\beta = 1$, however, we also consider the dependence of our results on $\beta$ below.

We derive the flow equation for general shape functions $R_k^g (z)$ and arbitrary endomorphism parameters $\beta$. To solve the threshold functions in the resulting $\beta$-functions explicitly, we mostly use the Litim cutoff \cite{Litim:2001up} with a prefactor $\alpha$,
\begin{equation}
    R_k^g (z) = \frac{\alpha k^2}{32 \pi G_N} \left(1 - \frac{z}{k^2}\right) \theta \left(1 - \frac{z}{k^2}\right) \, .
    \label{eqn:regdef}
\end{equation}
While analytical expressions for the threshold functions are obtained for general $\alpha$, they take a particularly simple form for $\alpha = 1$.

A second shape function that still allows obtaining analytical results for all threshold functions is the ``square'' of the Litim cutoff,
\begin{equation}
    R_k^g (z) = \frac{\alpha k^2}{32 \pi G_N} \left(1 - \frac{z}{k^2}\right)^2 \theta \left(1 - \frac{z}{k^2}\right) \, ,
    \label{eqn:reg2def}
\end{equation}
which we also refer to as Litim-2 cutoff. This additional shape function allows us to study some of the dependence of our results on the shape function.

\section{Results}\label{sec:results}

In this section, we present and discuss our findings for quantum gravity at sixth order in the derivative expansion of the \ac{MES}. The flow equation underlying our results has been obtained by an implementation of the setup outlined above in a computer algebra system. To ensure its correctness, two completely independent technical setups have been created based on (i) the \textsc{xAct} package for \textsc{Mathematica} \cite{Martin-Garcia:2007bqa,Brizuela:2008ra,Martin-Garcia:2008ysv,Nutma:2013zea} as implemented previously in the context of Asymptotic Safety \cite{Knorr:2021slg, Knorr:2021lll, Knorr:2022ilz}, and (ii) self-written \textsc{Mathematica} code together with \textsc{Form} \cite{Vermaseren:2000nd,Kuipers:2012rf}. This allows for important cross-checks at this order of the derivative expansion, where previously only very limited partial results have been available.\footnote{Where a comparison was possible, we completely agree with \cite{Knorr:2023usb}.} We provide the final result for all $\beta$- and $\gamma$-functions for arbitrary shape functions in a Mathematica notebook \cite{SupplementalMaterial}.

First, we report our results for fixed technical parameters $\alpha$ and $\beta$ and discuss the resulting \ac{UV} fixed point and its \ac{RG} flow towards the \ac{IR}. Following that, we also discuss the dependence of our results on $\alpha$ and $\beta$ and analyse the stability of the fixed point.

\subsection{UV Fixed Point}\label{sec:uvfp}

In this subsection, we choose fixed values for the technical parameters of our setup and search for non-trivial \ac{UV} fixed points. We fix a type II cutoff with the natural endomorphism ($\beta = 1$), while the shape function of the regulator is set to the Litim cutoff with $\alpha = 1$. To search for non-trivial fixed points, we define dimensionless couplings $g_{N}$ and $g_{C^3}$ via
\begin{equation}
    g_N = k^2 G_N \, , \qquad g_{C^3} = k^2 G_{C^3} \, ,
\end{equation}
as well as dimensionless $\gamma$-functions
\begin{equation}
    \dimlessgamma_X = k^{-d_X} \, \gamma_X \, ,
\end{equation}
where $d_X$ is the mass dimension of $\gamma_X$.
To find the value for the vacuum energy $\rho$ which we fix along the \ac{RG} flow, we set all $\gamma$-functions to zero and send $g_N \to 0$. At this point we look for a solution to the flow of $ \rho  =  \tilde{\rho} \, k^4$.
For $\alpha = 1$ and the Litim cutoff, we find for the vacuum energy\footnote{Note that the vacuum energy depends on $\alpha$ as well as the shape function of the cutoff, but not on $\beta$.}
\begin{equation}
    \rho = \frac{k^4}{8 \pi} \, .
\end{equation}
In the following, fixed point values are indicated by asterisks.

To investigate convergence properties of our system, we evaluate the flow order by order in the derivative expansion. At orders $\order{\partial^2}$ and $\order{\partial^4}$, Newton's coupling is the only dynamical coupling in the \ac{MES}. The fixed point results with a type I cutoff $(\beta = 0)$ were first obtained in \cite{Baldazzi:2021ydj}. With the type II cutoff, these results are qualitatively unchanged, see also \cite{Knorr:2022ilz}. At leading order $\order{\partial^2}$ and with the type II cutoff, the fixed point for Newton's coupling and its corresponding critical exponent read
\begin{equation}
    g_N^\ast = 0.368 \, , \qquad \theta = 2.239 \, .
    \label{eqn:EHres}
\end{equation}
At this order, the \ac{RG} kernel is parameterised by a single $\gamma$-function that takes the value
\begin{equation}
    \dimlessgamma_g^\ast = -0.987 \, .
\end{equation}
Due to the Gauss-Bonnet term being topological, the fixed point at order $\order{\partial^4}$ is still described by a single dynamical coupling while the \ac{RG} kernel contains three $\gamma$-functions,
\begin{gather}
    g_N^\ast = 0.364 \, , \quad \theta = 2.228 \, , \label{eqn:Quadres} \\
    \dimlessgamma_g^\ast = -0.998 \, , \quad \dimlessgamma_R^\ast = 0.014 \, , \quad \dimlessgamma_S^\ast = 0.057 \, .
\end{gather}
For a discussion of the topological term, we refer to appendix \ref{app:GB}.

A second essential and dynamical coupling is encountered at order $\order{\partial^6}$. This is the Goroff-Sagnotti coupling $g_{C^3}$. However, its inclusion does not lead to any qualitative changes of the fixed point. There is still a unique non-trivial fixed point at this order connected to the Gaussian fixed point. Its coupling coordinates are given by
\begin{equation}
 g^\ast_N = 0.364, \qquad g_{C^3}^\ast = 4.490 \cdot 10^{-7} \, .
 \label{eqn:fpcoup}
\end{equation}
Note the numerically small value of the Goroff-Sagnotti coupling. The critical exponents of the fixed point are
\begin{equation}
 \theta_1 = 2.225, \qquad \theta_2 = -3.850 \, .
\end{equation}
The non-trivial fixed point only admits one relevant direction -- the classically irrelevant Goroff-Sagnotti term stays irrelevant, confirming the findings in the standard scheme \cite{Gies:2016con}. This means that, after fixing the \ac{RG} scale, no free parameters are left in the theory.

{\renewcommand{\arraystretch}{1.3}
\begin{table}
\begin{tabular}{c|c|c}
  ~~~~~~order~~~~~~ & ~~~~~~$\theta_1$~~~~~~ & ~~~~~~$\theta_2$~~~~~~ \\ \hline
  $\mathcal O(\partial^2)$ & $2.239$ & --- \\
  $\mathcal O(\partial^4)$ & $2.228$ & --- \\
  $\mathcal O(\partial^6)$ & $2.225$ & $-3.850$ \\
  $\text{PMS at }\mathcal O(\partial^6)$ & $2.145$ & $-2.697$
\end{tabular}
\caption{\label{tab:convergence}Comparison of critical exponents at different orders of the derivative expansion, using $\alpha=\beta=1$ and the Litim shape function. The last line gives \ac{PMS} values obtained in \eqref{eqn:PMSLitim1} for the critical exponents at $\mathcal O(\partial^6)$.}
\end{table}
}

Compared to the lower order results, \eqref{eqn:EHres} and \eqref{eqn:Quadres}, the fixed point value of Newton's coupling only receives minor corrections at order $\order{\partial^6}$.
This is also true for the critical exponents that we compare in \autoref{tab:convergence}. The first two significant figures of $\theta_1$ do not change between the leading order and $\order{\partial^6}$. The relative change between $\order{\partial^4}$ and $\order{\partial^6}$ is as small as $0.1\%$. This behaviour of the derivative expansion suggests convergence towards higher orders. Moreover, we note that the value of the critical exponent $\theta_1$ is close to its canonical value $\theta_1^\text{cl} = 2$, achieved in a one-loop approximation. This means that the induced quantum correction for $\theta_1$ is small.

The stability of the fixed point in the \ac{MES} can be quantitatively understood by the \ac{RG} kernel \eqref{eqn:RGkernel}. At the fixed point, we find the following values for the $\gamma$-functions:
\begin{equation}
\begin{aligned}
 \dimlessgamma_g^\ast &= -0.996,& \, \dimlessgamma_R^\ast &= 0.011, \\
 \dimlessgamma_S^\ast &= 0.056,& \, \dimlessgamma_{R^2}^\ast &= 0.008, \\ 
 \dimlessgamma_{C^2}^\ast &= -0.005,& \, \dimlessgamma_{RS}^\ast &= -0.012, \\
 \dimlessgamma_{SSTL}^\ast &= -0.024,& \, \dimlessgamma_{CS}^\ast &= 0.025, \\
 \dimlessgamma_{\Delta R}^\ast &= -0.011,& \, \dimlessgamma_{\Delta S}^\ast &= 0.007 \, .
\end{aligned}
\label{eqn:gammas}
\end{equation}
Except for $\dimlessgamma_g$, which is already included in the Einstein-Hilbert approximation, the absolute values of all of the $\gamma$-functions are much smaller than one.
Together with the small value of the Goroff-Sagnotti coupling at the fixed point, this quantitatively explains why higher order effects on the fixed point are small. 

To close this subsection, let us note that the non-trivial fixed point shows several signs of a near-perturbative behaviour. This includes the fixed point value of the Goroff-Sagnotti coupling, quantum corrections to the critical exponents, and the $\gamma$-functions which parameterise the \ac{RG} kernel. All of these take very small values within the \ac{MES}. The smallness of the \ac{RG} kernel also implies that the structure of the redundant operators is a small deformation of the Gaussian fixed point. Together with earlier indications of a near-perturbative behaviour of quantum gravity \cite{Falls:2013bv,Falls:2014tra, Eichhorn:2018akn, Eichhorn:2018ydy}, our results strengthen the evidence that the non-trivial fixed point is in a near-perturbative region. As a matter of fact, if we were to neglect all $\gamma$-functions except $\gamma_g$ in the above computation, the results change only insignificantly. In this reduced setup, we find
\begin{equation}
    g^\ast_N = 0.368 \, , \, \, \, g_{C^3}^\ast = 7.704 \cdot 10^{-7} \, , \, \, \, \dimlessgamma_g^\ast = -0.987 \, ,
\end{equation}
and
\begin{equation}
    \theta_1 = 2.239 \, , \qquad \theta_2 = -3.881 \, .
    \label{eqn:simple}
\end{equation}
This simplified approximation compares very favourably with the full computation, and opens up a path forward to high order computations: keep all essential couplings and $\gamma_g$, while neglecting all other $\gamma$-functions. It is curious that the same procedure in the standard scheme yields a highly non-canonical critical exponent \cite{Gies:2016con}, see also appendix \ref{app:C3standard}.

\subsection{RG Flow and IR Behaviour}\label{sec:RGglowandIR}

\begin{figure}
	\includegraphics[width=\columnwidth]{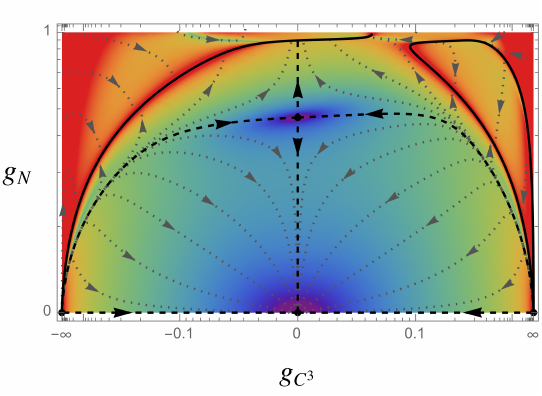}
	\caption{\label{fig:PD}Phase diagram with the Litim shape function and technical parameters $\alpha = \beta = 1$. Arrows point from the \ac{UV} to the \ac{IR}, and the colour gradient indicates the flow velocity (increasing velocity from blue to red). Solid black lines indicate divergences of $\beta$-functions, dashed black lines represent separatrices, whereas dotted grey lines are sample trajectories.}
\end{figure}

To visualise the flow of asymptotically safe trajectories and their \ac{IR} properties, \autoref{fig:PD} shows the phase diagram within the \ac{MES} for $\alpha = \beta = 1$ and Litim cutoff.
Since we require Newton's coupling to be small and positive at low energies, any physical \ac{RG} trajectory must reach the vicinity of the Gaussian fixed point with a positive Newton coupling in the \ac{IR}. How couplings arrive at the  Gaussian fixed point from the \ac{UV} can be read off from \autoref{fig:PD}. Among these trajectories that have an \ac{IR} limit corresponding to perturbative \ac{GR}, there are two possibilities for their \ac{UV} origin.

The first and generic option consists of trajectories for which the Goroff-Sagnotti coupling diverges at high energies. These trajectories begin in the UV at $g_N = 0$ and $g_{C^3} = \pm \infty$ and therefore do not correspond to asymptotically safe theories. Based on scaling arguments \cite{Weinberg:1980gg}, one expects that observables such as scattering amplitudes have a divergent high energy behaviour.

Finite values for the couplings are only reached on the separatrix between the regions where the Goroff-Sagnotti coupling grows towards $+ \infty$ and $-\infty$. This is the asymptotically safe trajectory which originates from the fixed point \eqref{eqn:fpcoup} in the \ac{UV}, and it is the only trajectory that has a well-defined behaviour at all energy scales and connects to the region of perturbative quantum \ac{GR} at low energy scales. There are no free parameters that need to be fixed and no second non-trivial fixed point is available as a \ac{UV} completion for trajectories that are connected to the Gaussian fixed point.

\begin{figure*}
    \centering
    \includegraphics[width=0.32\textwidth]{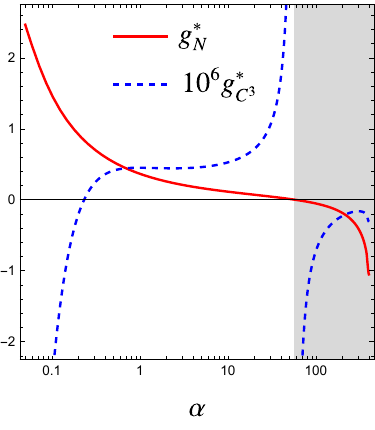}
    \includegraphics[width=0.32\textwidth]{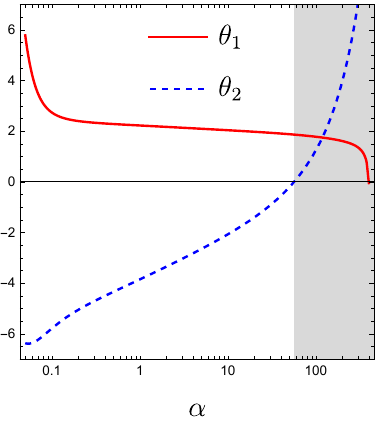}
    \includegraphics[width=0.32\textwidth]{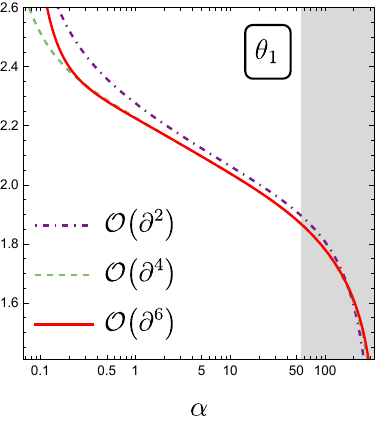}
    \caption{\label{fig:alphadepfp}Fixed point dependence on $\alpha$ with the type II cutoff $(\beta = 1)$ and the Litim cutoff. The left plot shows the couplings of the fixed point at order $\order{\partial^6}$ in the derivative expansion. The central plot shows the critical exponents at order $\order{\partial^6}$. The right panel compares the $\alpha$-dependence of the relevant critical exponent obtained at orders $\order{\partial^2}$, $\order{\partial^4}$, and $\order{\partial^6}$. In all plots, the grey-shaded area indicates regions of negative Newton's coupling.}
\end{figure*}

The \ac{UV} safe separatrix arises from the relevant deformation of the \ac{UV} fixed point. The irrelevant deformation spans separatrices which split the phase diagram into two parts: 
all trajectories with smaller Newton coupling than these separatrices reach the Gaussian fixed point in the \ac{IR}, while trajectories with larger Newton coupling do not. The latter eventually end up in a divergence of the \ac{RG} flow. 
Such divergences arise when denominators of the $\beta$-functions develop a zero. The zeros of those denominators form hypersurfaces in coupling space at which the $\beta$-functions diverge. In \autoref{fig:PD} we see that this happens whenever Newton's coupling becomes to large, and we attribute these to a breakdown of our approximation.

The asymptotically safe trajectory that connects the non-trivial fixed point in the \ac{UV} with the Gaussian fixed point in the \ac{IR} leads to a non-trivial prediction at low energies. For small $k$, we can solve the $k$-dependence of the Goroff-Sagnotti coupling in terms of Newton's coupling,
\begin{equation}\label{eq:GSIR}
    g_{C^3} = \left[G_N k^2 \right] \left( A - \frac{43}{645120\pi^3} \ln G_N k^2 \right) \, .
\end{equation}
The prefactor of the logarithm is identical for all trajectories reaching the Gaussian fixed point. By contrast, the Wilson coefficient $A$ depends on the trajectory. For the asymptotically safe trajectory, the value of $A$ is predicted by the fixed point. Numerically, we find
\begin{equation}\label{eq:GSWilsonCoefficient}
 A = -3.988 \cdot 10^{-6} \, ,
\end{equation}
once again for the choice $\alpha=\beta=1$ and the Litim shape function.

In principle, the \ac{RG} running in the \ac{IR} can be compared to perturbative $\beta$-functions. On the perturbative side, additional attention must be paid to the fact that the Gauss-Bonnet term acts as an evanescent operator which leads to a non-trivial relation between the perturbative two-loop divergence and its \ac{RG} running \cite{Bern:2015xsa,Bern:2017puu}. Taking this into account, the perturbative $\beta$-function of the Goroff-Sagnotti coupling is given by
\begin{equation}
    \partial_t g_{C^3} = 2 g_{C^3} + \frac{1}{7680 \pi^3} \, g_N + \dots \, .
    \label{eqn:GSpert}
\end{equation}
Our result using the type II cutoff is
\begin{equation}
    \partial_t g_{C^3} = 2 g_{C^3} - \frac{43}{322560 \pi^3} \, g_N + \dots \, .
    \label{eqn:GSfRGpert}
\end{equation}
Although the two results for the quantum corrections do not agree, it should be noted that beyond the one-loop order, a comparison with perturbative results requires approximation schemes other than a derivative expansion in the \ac{FRG}. Indeed, in \cite{Papenbrock:1994kf,Baldazzi_2020} it is shown that two-loop results for the quartic coupling of an $O(N)$ symmetric scalar field can be obtained with an infinite number of terms in the derivative expansion. Thus, it is reasonable to think that also the case of gravity requires an infinite expansion in derivatives to reproduce the perturbative result \eqref{eqn:GSpert}.

\subsection{Regulator Dependence with Type II Cutoff}\label{sec:regdep}

\begin{figure*}
    \centering
    \includegraphics[width=0.65\textwidth]{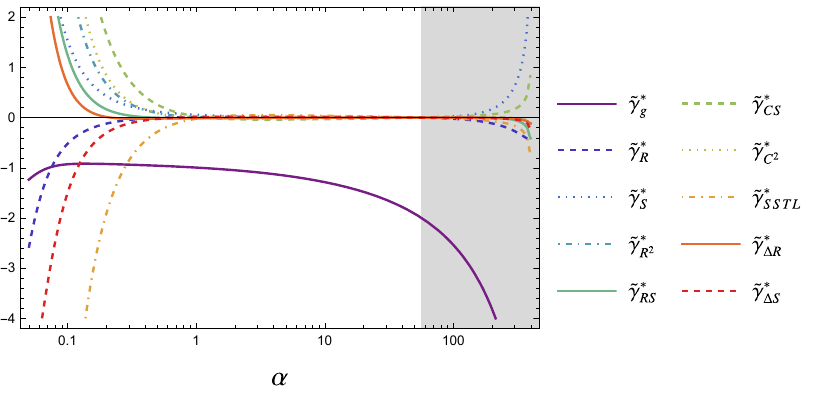}
    \caption{The $\gamma$-functions are shown in dependence of $\alpha$ for the type II cutoff ($\beta = 1$) and the Litim cutoff. In the grey shaded area, Newton's coupling takes negative values.}
    \label{fig:alphadepgamma}
\end{figure*}

To test the reliability of our results, we analyse the regulator dependence. For this purpose, the parameters $\alpha$ and $\beta$ are introduced in the regulator. In this subsection, we focus on the dependence on one of these parameters while keeping the other fixed: we continue using the type II cutoff with $\beta = 1$, while changing the strength of the regulator by varying $\alpha$.

In \autoref{fig:alphadepfp}, we show the dependence of the fixed point on $\alpha$. There is a large area in which the fixed point is stable. This reaches from $\alpha \approx 0.05$ to $\alpha \approx 50$. Going beyond this regime, the couplings and critical exponents grow large. Furthermore, the fixed point becomes unphysical due to a sign change of Newton's coupling at $\alpha \approx 57$. For $\alpha > 57$, the non-trivial fixed point is located at negative Newton coupling. Due to zeros and divergences of the $\beta$-functions at $g_N = 0$ and at large $g_N$, such a fixed point cannot admit an \ac{RG} trajectory connecting it to a low energy regime with small positive Newton's coupling. 
Moreover, the sign change of Newton's coupling at $\alpha \approx 57$ is accompanied by a divergence of the Goroff-Sagnotti coupling.
Thus, we expect that the physical region in which our approximation scheme is well-defined does not reach beyond $\alpha = 57$.

Similar observations hold for the small $\alpha$ regime. Although we do not observe a divergence in the couplings, the critical exponents diverge for $\alpha \to 0$. This is indicated in \autoref{fig:alphadepfp} which shows how the relevant critical exponent becomes large for $\alpha < 0.1$. 
The limit $\alpha \to 0$ is typically characterized by log-tails in $\alpha$ for even dimensions: in \cite{Baldazzi_2021}, this has been discussed for a scalar field in two and four dimensions. 

In the intermediate range, $0.1 < \alpha < 57$, we observe a large area of stability for the fixed point. Most notably, this includes a plateau for the relevant critical exponent. From $\alpha \approx 0.2$ up to $\alpha \approx 50$, we find a mild dependence of $\theta_1$ on $\alpha$, taking values from $2.4$ to $1.9$.
This is not only true at order $\order{\partial^6}$, but also at lower orders. In fact, the relevant critical exponent only receives minor numerical modifications between different orders from $\order{\partial^2}$ up to $\order{\partial^6}$ throughout the whole region $0.1 < \alpha < 57$. Starting from $\alpha = 0.3$ and up until $\alpha = 57$, the difference between $\order{\partial^4}$ and $\order{\partial^6}$ is of the order of $0.1\%$.

This observed convergence between different orders of the derivative expansion can, again, be explained by the smallness of the Goroff-Sagnotti coupling and the \ac{RG} kernel. In \autoref{fig:alphadepgamma}, we show the $\gamma$-functions of the \ac{RG} kernel at the fixed point in dependence of $\alpha$. Within the previously identified region, all $\gamma$-functions except for $\gamma_g$ are essentially vanishing. Thus, different orders in the derivative expansion are almost disentangled from each other. Higher orders of the derivative expansion only have minimal effects on results obtained at lower orders. Therefore, we expect our results to be robust against effects from higher order curvature contributions.

\begin{figure*}
    \includegraphics[width=\columnwidth]{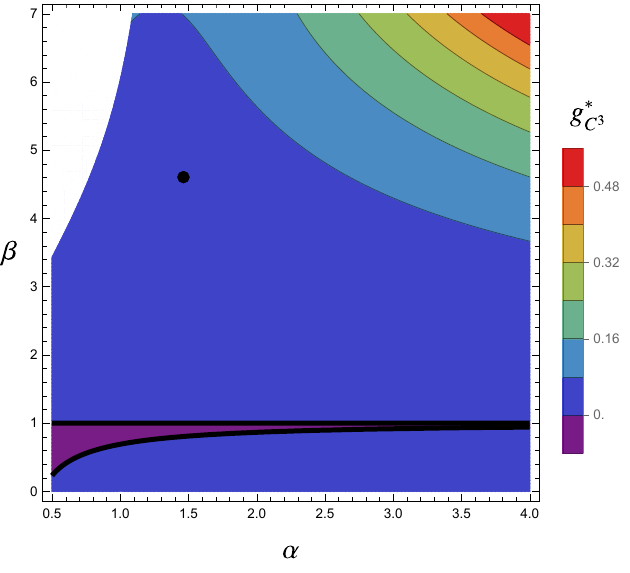}
    \hfill
    \includegraphics[width=0.49\textwidth]{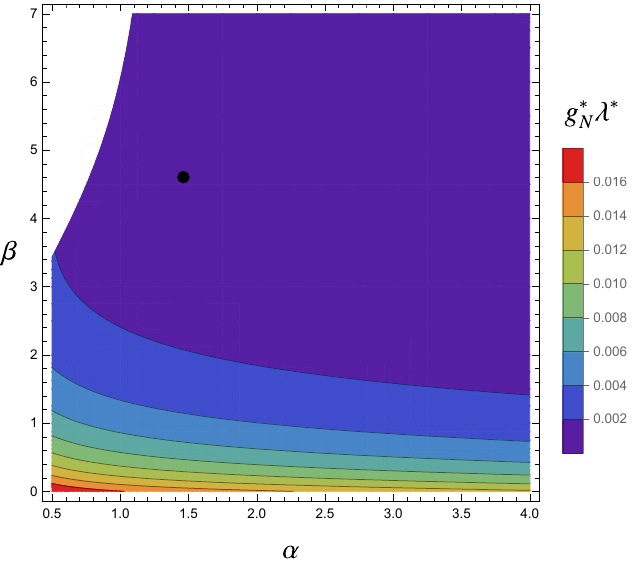}
    \caption{The Goroff-Sagnotti coupling (left panel) and the product of Newton's coupling and the cosmological constant (right panel) in dependence of $\alpha$ and $\beta$ with the Litim cutoff. Thick black lines indicate zeros of the Goroff-Sagnotti coupling. Note the pronounced zero along $\beta = 1$ corresponding to the type II cutoff. Similar zeros are also seen for the $\gamma$-functions that parameterise the \ac{RG} kernel, see \autoref{fig:gammafuncs}. Similar results are found for the Litim-$2$ cutoff.}
    \label{fig:gC3}
\end{figure*}

Following this discussion, it is worth noting that the $\alpha$-dependence of the relevant critical exponent is virtually unaffected by higher order operators beyond four powers of derivatives, see right panel of \autoref{fig:alphadepfp}. This behaviour is different compared to other models that have been studied in the derivative expansion such as the Ising model \cite{Balog:2019rrg}. There, the dependence on $\alpha$ usually becomes stronger with increasing orders. This signals that the derivative expansion only converges for specific values of $\alpha$. In contrast, \autoref{fig:alphadepfp} suggests a large range for $\alpha$ in which the $\alpha$-dependence does not change notably beyond $\order{\partial^4}$. This could point towards a finite $\alpha$-range in which the derivative expansion converges. 
However, the $\alpha$-dependence does not become flat in this regime. Most likely, this can be explained by truncation artefacts of the background field approximation. The $\alpha$-dependence would then be caused by neglecting non-diffeomorphism invariant terms in the action. The inclusion of those would potentially level out the $\alpha$-dependence within the domain of convergence. To test this hypothesis and control this effect, it is necessary to go beyond the background field approximation.

Compared to the relevant critical exponent $\theta_1$, the critical exponent of the Goroff-Sagnotti coupling $\theta_2$ shows a stronger dependence on $\alpha$. This is not surprising since the Goroff-Sagnotti term is included for the first time in the derivative expansion at order $\order{\partial^6}$. Most likely, its critical exponent is stabilised by other operators at higher orders. From $\alpha = 0.2$ to $\alpha = 50$, the critical exponent $\theta_2$ takes values from $\theta_2 = -5$ to $\theta_2 = -0.2$. It is decreasing in magnitude with increasing $\alpha$. At $\alpha \approx 57$ it changes sign and becomes relevant. However, this is the point where Newton's coupling becomes negative and falls out of the physical regime. Thus, within our approximation the Goroff-Sagnotti term stays irrelevant within the physical region.

Let us note that an analysis of the $\alpha$-dependence of the fixed point shows that previously observed indications of near-perturbativity are not an artifact of a specific value for $\alpha$. Instead, the smallness of the Goroff-Sagnotti coupling, the $\gamma$-functions, and corrections to the critical exponents is independent of $\alpha$ within the region in which we expect our approximation to be trustworthy. Hence, we tend to conclude that the non-trivial fixed point is near-perturbative, independently of $\alpha$.

\subsection{Regulator Dependence Beyond Type II Cutoff}\label{sec:regdepbeyond}

We now analyse the full regulator dependence of our results, varying both $\alpha$ and $\beta$. Thus, in addition to the regulator strength encoded in $\alpha$, we test the dependence of our results on the endomorphism parameter $\beta$. For $\beta = 1$, the regulator corresponds to a type II cutoff which has been studied in detail above. For $\beta = 0$, we recover the type I cutoff. We have computed the fixed point for all values of $0.5 < \alpha < 4$, and $0 < \beta < 7$, using the two cutoff shapes given in \eqref{eqn:regdef} and \eqref{eqn:reg2def}.

\begin{figure*}
    \centering
    \quad \ \, \includegraphics[width=0.32\textwidth]{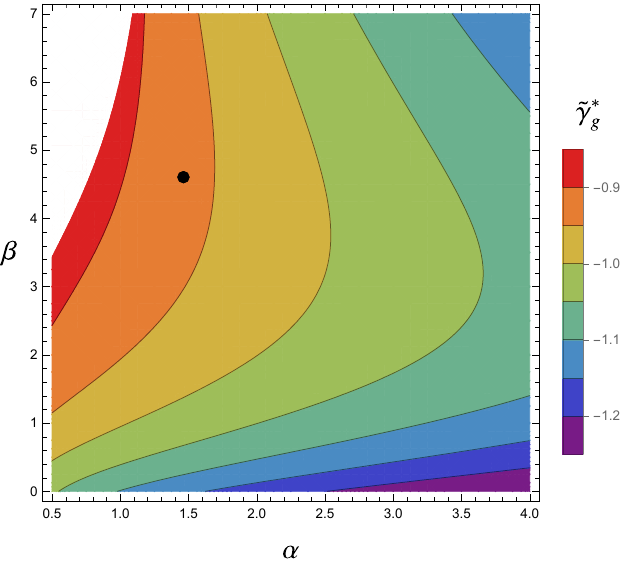}
    \\
    \includegraphics[width=0.32\textwidth]{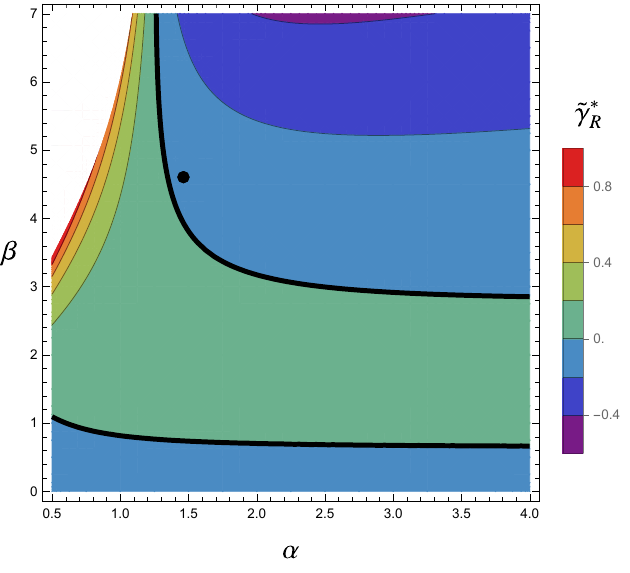}
    \hfill
    \includegraphics[width=0.32\textwidth]{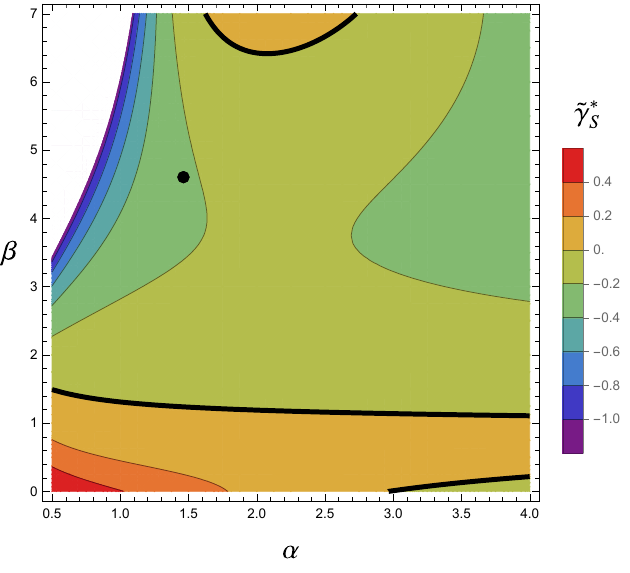}
    \includegraphics[width=0.32\textwidth]{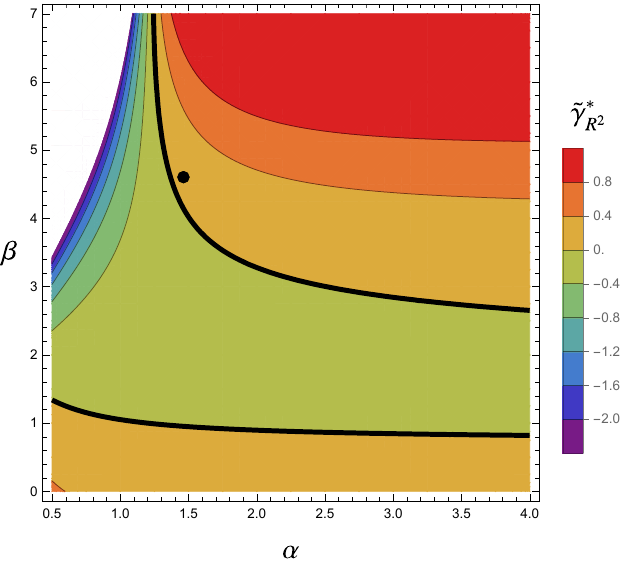}
    \hfill
    \includegraphics[width=0.32\textwidth]{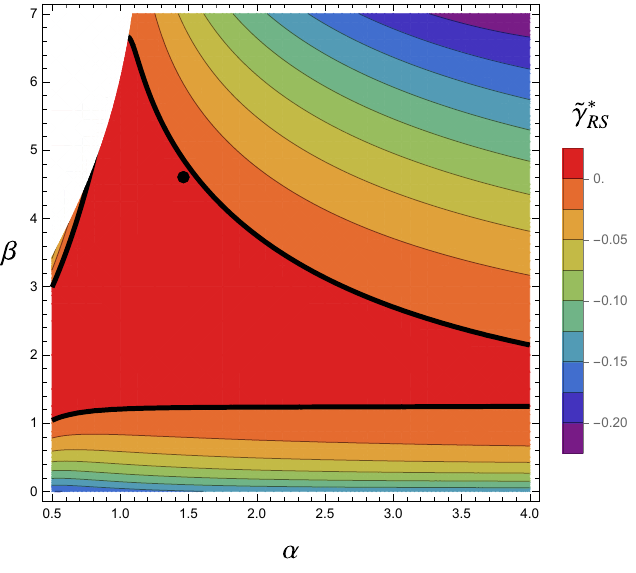}
    \hfill
    \includegraphics[width=0.32\textwidth]{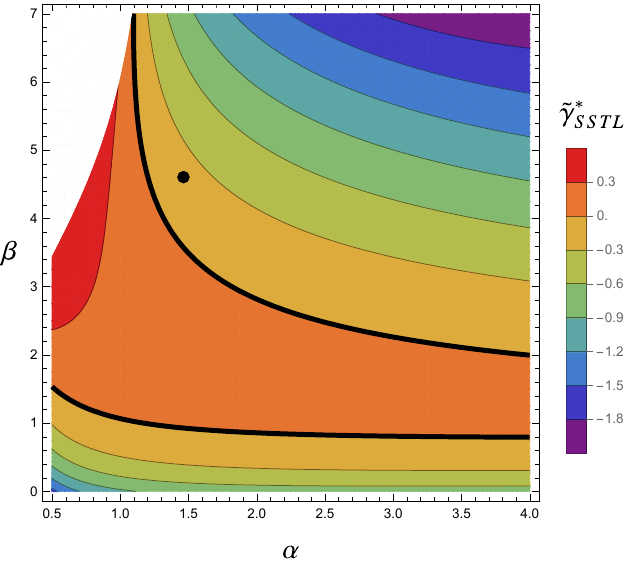}
    \includegraphics[width=0.32\textwidth]{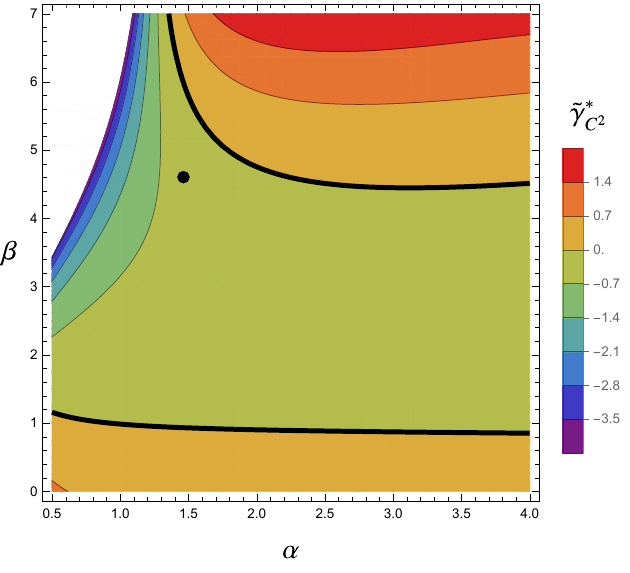}
    \hfill
    \includegraphics[width=0.32\textwidth]{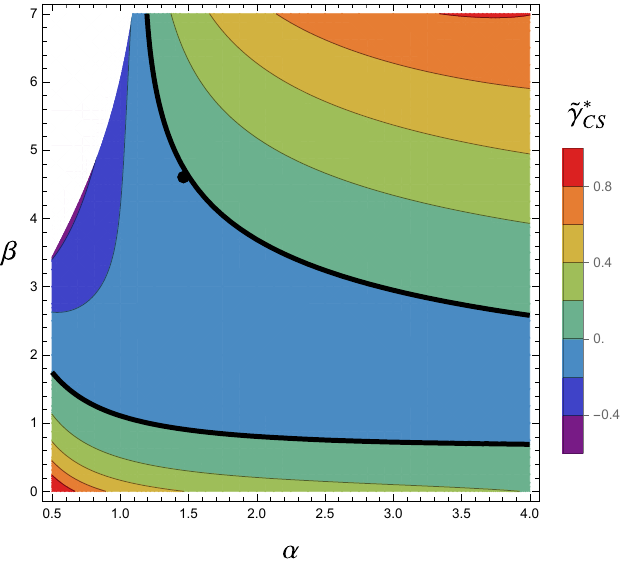}
    \hfill
    \includegraphics[width=0.32\textwidth]{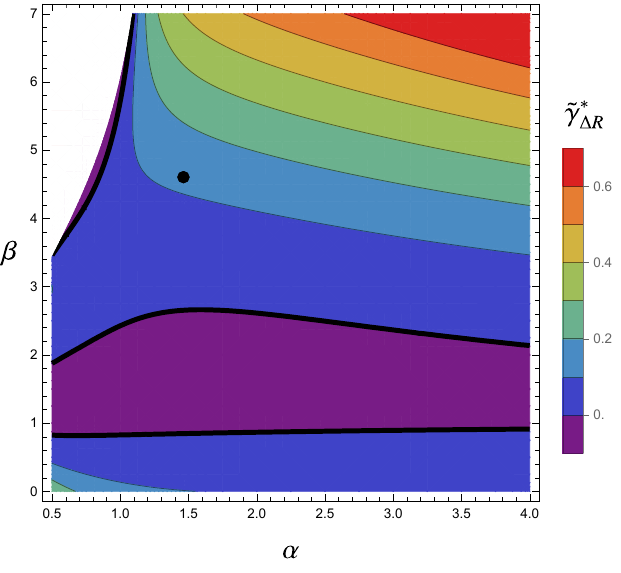}
    \includegraphics[width=0.32\textwidth]{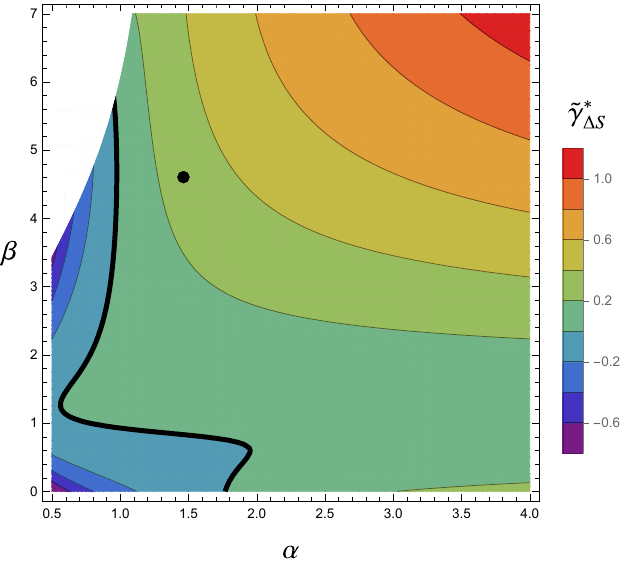}
    \caption{\label{fig:gammafuncs}Values for the $\gamma$-function at the fixed point in dependence of $\alpha$ and $\beta$ with the Litim cutoff. Zeros are indicated by thick black lines.}
\end{figure*}

Apart from parameter choices for which the fixed point becomes unstable, our qualitative results are independent of $\alpha$ and $\beta$. Within the scanned region, instabilities only occur for small $\alpha$ and large $\beta$. In this area, the fixed point moves close to a hypersurface on which the \ac{RG} flow develops a singularity. This leads to rapidly changing and large values for the fixed point couplings, its critical exponents, and the \ac{RG} kernel. As a consequence, we cannot find the fixed point reliably anymore, and we expect that our approximation scheme cannot be trusted in this regime. Accordingly, this regime is excluded from all plots.

One of the main results of the scan in the $(\alpha,\beta)$-plane is the observed special role of $\beta = 1$. As can be seen in \autoref{fig:gC3}, the Goroff-Sagnotti coupling is zero at the fixed point almost exactly along the line of $\beta = 1$, while taking comparatively large positive values for other values of $\beta$. Thus, the smallness of the Goroff-Sagnotti coupling is tied to $\beta = 1$. A similar effect is observed for the \ac{RG} kernel in \autoref{fig:gammafuncs}. Except for $\gamma_g$, all $\gamma$-functions develop zeros close to $\beta = 1$. These observations motivate our usage of the type II cutoff $(\beta = 1)$ above: it is this technical choice that makes the near-perturbativity of the non-trivial fixed point explicit, and leads to a small Goroff-Sagnotti coupling and \ac{RG} kernel, with the exception of $\gamma_g$. We might expect our approximation to converge most quickly in this regime as higher order curvatures are as disentangled as possible from lower order curvatures.

\begin{figure*}
    \centering
    \begin{minipage}[t]{0.49\textwidth}
        \large Litim cutoff \\[1ex]
        \includegraphics[width=\textwidth]{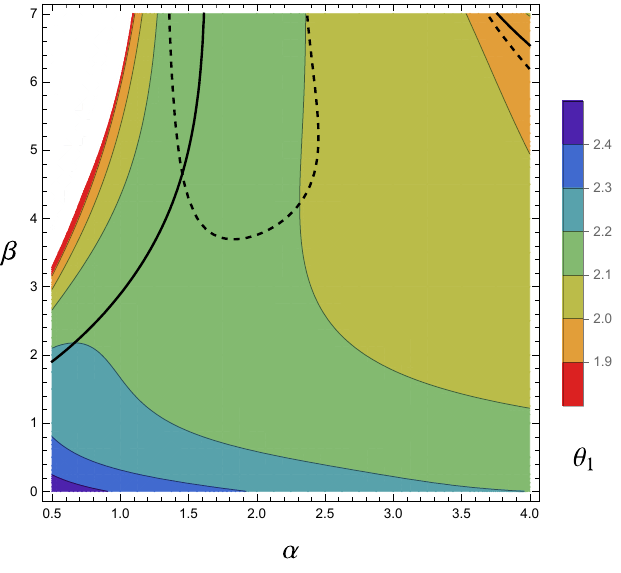}
        \includegraphics[width=\textwidth]{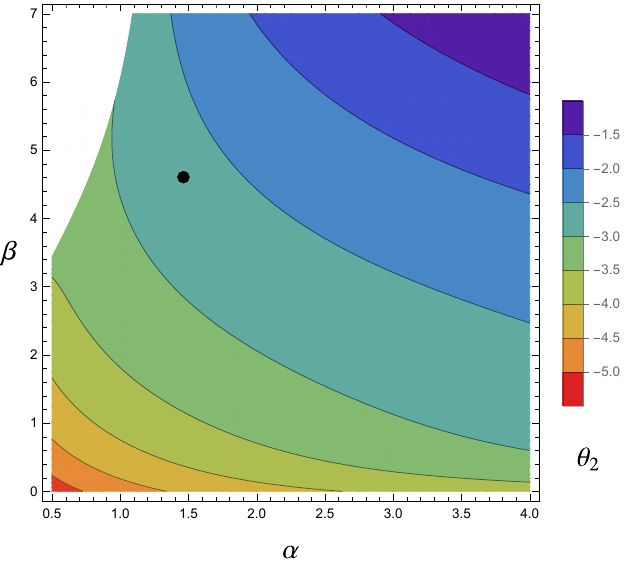}
    \end{minipage}
    \hfill
    \begin{minipage}[t]{0.49\textwidth}
        \large Litim-$2$ cutoff \\[1ex]
        \includegraphics[width=\textwidth]{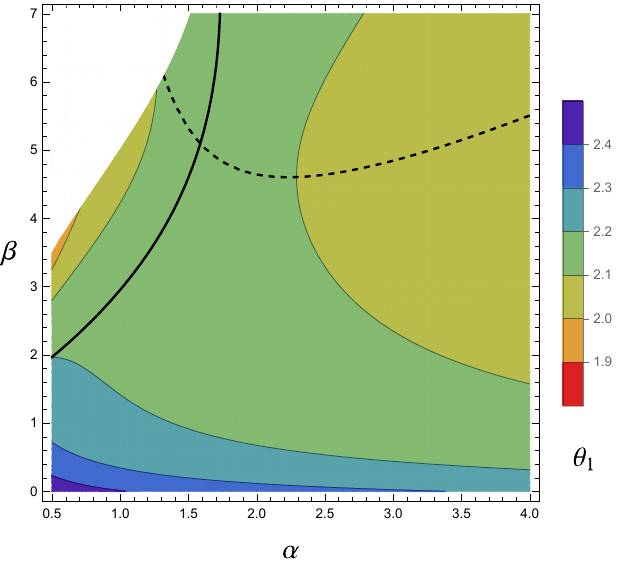}
        \includegraphics[width=\textwidth]{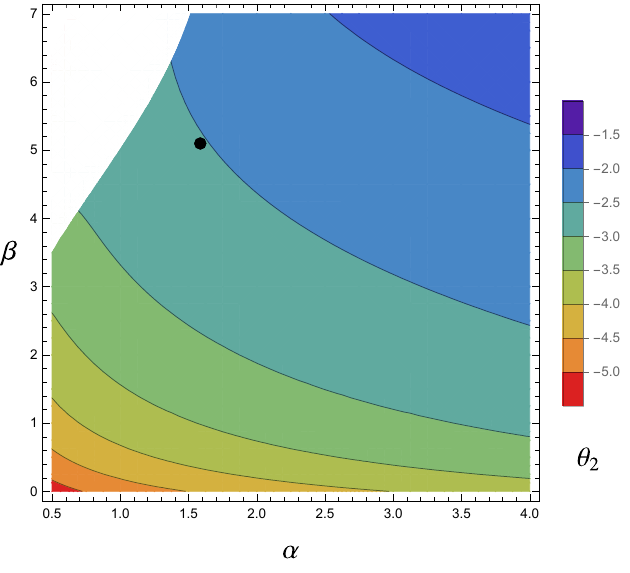}
    \end{minipage}
    \caption{Regulator dependence of the critical exponents on $\alpha$ and $\beta$ with Litim (left column) and Litim-2 (right column) cutoff. Thick solid (dashed) black curves show where the dependence of the relevant exponent on $\alpha$ ($\beta$) vanishes. A black dot denotes the \ac{PMS} value. For small $\alpha$ and large $\beta$, we find that the $\gamma$-functions become large, and critical exponents fluctuate strongly. This unstable area has been cut out from the plots.}
    \label{fig:PMSRel}
\end{figure*}

In \autoref{fig:PMSRel}, we show the dependence of the relevant and irrelevant critical exponent on $\alpha$ and $\beta$, using the Litim as well as the Litim-$2$ cutoff. With both cutoffs, the relevant critical exponent takes values between $2.0$ and $2.2$ in almost the entire scanned area. This includes most of the $\beta = 1$ line, and shows that the relevant critical exponent is also stable away from $\beta = 1$. Towards smaller values of $\alpha$, the fluctuations of the relevant critical exponent become stronger. This is in line with our results for the type II cutoff in \autoref{sec:regdep} and similar studies in scalar field theories \cite{Baldazzi_2021}. Compared to the relevant critical exponent, the irrelevant critical exponent shows stronger variations. Within the scanned region, these are almost uniform and take values $-1.5$ and $-5$. Thus, the irrelevant critical exponent is strictly irrelevant.

We can also apply the \acf{PMS} to determine preferred values for $\alpha$ and $\beta$ for which we expect the derivative expansion to rapidly converge. This has proven successful in determining the critical exponents of the Ising model \cite{Balog:2019rrg}. The \ac{PMS} is based on the fact that physical observables will not depend on the regulator at the exact level. In approximations, this will not be the case but we can impose it as an extra restriction in addition to the fixed point condition and therefore determine \ac{PMS} values for $\alpha$, $\beta$, the fixed point and the critical exponents.
We therefore consider the \ac{PMS}, and determine technical parameters for which the dependence of physical observables on $\alpha$ and $\beta$ vanishes. This is possible if an observable obtains an extremum as a function of technical parameters. Applying this concept to the critical exponents, we do not find a \ac{PMS} point for $\alpha$ at $\beta = 1$. This is clear from \autoref{fig:alphadepfp} which shows that the critical exponents do not take an extremum anywhere within the physical regime for $\alpha$. However, a \ac{PMS} point does exist if we allow for $\beta \neq 1$ and require the relevant critical exponent $\theta_1$ to have an extremum in both $\alpha$ and $\beta$. In \autoref{fig:PMSRel}, solid black lines show where the dependence of the relevant critical exponent on $\alpha$ vanishes, while dashed black lines indicate where the dependence on $\beta$ vanishes. Within the scanned region, we find a unique intersection of both -- this holds for the Litim and the Litim-$2$ cutoff. For the Litim cutoff, the resulting \ac{PMS} value is located at
\begin{equation}
    \begin{aligned}
        \alpha &=\, 1.454 \, , & \beta &=\, 4.655 \, ,\\
        \theta_1 &=\, 2.145 \, , & \theta_2 &=\, -2.697 \, , \\
        g_N^\ast &=\, 0.108 \, , & g_{C^3}^\ast &=\, 0.034 \, , \\
        \dimlessgamma_g^\ast &=\, -0.933\,,& \dimlessgamma_R^\ast &=\, -0.036 \,,\\
        \dimlessgamma_S^\ast &=\, -0.226 \,,&  \dimlessgamma_{R^2}^\ast &=\, 0.095 \,,\\
        \dimlessgamma_{RS}^\ast &=\, 0.004 \,,& \dimlessgamma_{CS}^\ast &=\, -0.011 \,,\\
        \dimlessgamma_{C^2}^\ast &=\, -0.430 \,,& \dimlessgamma_{SSTL}^\ast &=\, -0.189 \,,\\
        \dimlessgamma_{\Delta R}^\ast &=\, 0.125 \,,& \dimlessgamma_{\Delta S}^\ast &=\, 0.280\,,
    \end{aligned}
    \label{eqn:PMSLitim1}
\end{equation}
while we find for the Litim-$2$ cutoff,
\begin{equation}
    \begin{aligned}
        \alpha &=\, 1.586 \,,& \beta &=\, 5.098 \,,\\
        \theta_1 &=\, 2.117 \,,& \theta_2 &=\, - 2.521 \,,\\
        g_N^\ast &=\, 0.119 \,,& g_{C^3}^\ast &=\, 0.101 \,,\\
        \dimlessgamma_g^\ast &=\, -0.894 \,& \dimlessgamma_R^\ast &=\, 0.061 \,,\\
        \dimlessgamma_S^\ast &=\, -0.787 \,,& \dimlessgamma_{R^2}^\ast &=\, -1.861 \,,\\
        \dimlessgamma_{RS}^\ast &=\, 0.735 \,,& \dimlessgamma_{CS}^\ast &=\, -0.865 \,,\\
        \dimlessgamma_{C^2}^\ast &=\, -5.071 \,,& \dimlessgamma_{SSTL}^\ast &=\,  1.626 \,,\\
        \dimlessgamma_{\Delta R}^\ast &=\, -0.107 \,,& \dimlessgamma_{\Delta S}^\ast &=\, 0.740 \,.
    \end{aligned}
    \label{eqn:PMSLitim2}
\end{equation}
Comparing both results, the relevant critical exponents differ by just $1 \%$, while the irrelevant exponents agree within $7\%$. Similar small differences arise when comparing to the results with the type II cutoff, \ie{} \autoref{tab:convergence}. The results for the relevant critical exponent agree within $5\%$. Stronger variations arise for the results of the irrelevant critical exponent, which shows larger dependencies on technical parameters, once again indicating that higher order operators are likely necessary to stabilise its value.

Similar to the type II cutoff, the \ac{PMS} point is an anchor point for zeros of most of the $\gamma$-functions, see \autoref{fig:gammafuncs}. Except for $\gamma_g$, $\gamma_S$, $\gamma_{\Delta R}$ and $\gamma_{\Delta S}$ all $\gamma$-functions develop a zero close to the \ac{PMS} point. The major difference compared with the type II cutoff at $\beta = 1$ is the value of the Goroff-Sagnotti coupling. While it vanishes close to $\beta = 1$, it does not have a zero in the vicinity of the \ac{PMS} point and takes a comparatively large value.

In summary, we have found that our qualitative -- and to a large extent quantitative -- results are independent of the technical parameters $\alpha$ and $\beta$. In particular, the critical exponent of the Goroff-Sagnotti coupling is always irrelevant. Moreover, the relevant critical exponent is remarkably stable, and takes values between $2.0 - 2.3$ within the trustworthy region of our setup. On the other hand, the smallness of the \ac{RG} kernel and the Goroff-Sagnotti coupling are to some extend related to special choices for technical parameters, such as the type II cutoff $(\beta = 1)$ or the \ac{PMS} point, see \eqref{eqn:PMSLitim1} and \eqref{eqn:PMSLitim2}. In both cases, the \ac{RG} kernel becomes small, while the former also leads to an extremely small Goroff-Sagnotti coupling. We expect that this near-perturbativity of couplings and \ac{RG} kernel is particularly useful for the convergence of our expansion and we expect the best convergence in this region.

\section{Discussion}\label{sec:discussion}

In this work, we have provided further evidence for the existence of a non-trivial fixed point in quantum gravity. Employing the \ac{FRG}, we have performed a consistent derivative expansion to sixth order on general backgrounds. This is the first complete work at this order in four dimensions,\footnote{For a recent discussion of one-loop $\beta$-functions of cubic gravity in its critical dimension, $d=6$, see \cite{Knorr:2021lll}.} and marks a considerable step forward compared to previous approaches that were limited to fourth order of the derivative expansion, or the use of simpler background geometries. Moreover, we have used recent insights on the use of field redefinitions in the \ac{FRG} by implementing the \ac{MES}. In this scheme, only essential couplings are retained in the action, while redundant couplings are fixed at the expense of non-trivial field redefinitions. In this way, it is possible to eliminate the inessential couplings from the flow equations, which then depends only on essential couplings. Two main advantages of this approach are (i) a simpler structure for the propagator, which makes computations to this order of the derivative expansion feasible, and (ii) focusing on the essential couplings of the theory when searching for non-trivial fixed points and evaluating their critical exponents. The latter point is important for the Asymptotic Safety scenario since redundant couplings do not affect physical observables, and consequently they should not be taken into account when counting the number of free parameters of the theory. Furthermore, the assumptions of the \ac{MES}, which identifies the inessential couplings at the free fixed point, restrict the theory space to the subspace where those couplings remain inessential. Namely, we focus on the universality classes connected to \ac{GR}, and characterized by the same redundant operator structure. The simple form of the propagator then suggests that fixed points found in the \ac{MES} have the same propagating degrees of freedom as \ac{GR}, as also indicated by recent results using momentum-dependent field redefinitions \cite{Knorr:2023usb}. More technically, on any conformally flat background, the propagator has the same pole structure as \ac{GR}.

Although a non-trivial gravitational fixed point has previously been identified at second and quartic orders in derivatives within the \ac{MES} \cite{Baldazzi:2021orb} at sixth order a second dynamical essential coupling enters which could have strongly altered results obtained at lower orders. Nevertheless, as seen in \autoref{fig:PD}, the results of our analysis suggest that the fixed point survives the inclusion of terms at sixth order in derivatives, and moreover, that it is remarkably stable. Newton's coupling and its critical exponent are essentially unaffected by the extension, even though a new essential coupling enters the system. Between fourth and sixth order of the derivative expansion, the relevant critical exponents receives corrections of only $\approx 0.1\%$, taking a value of $\theta_1 = 2.225$ at sixth order, see \autoref{tab:convergence}. Furthermore, the Goroff-Sagnotti term stays irrelevant at the fixed point, which confirms previous observations in the standard scheme \cite{Gies:2016con}. Therefore, its Wilson coefficient is a prediction of the theory, see \eqref{eq:GSIR} and \eqref{eq:GSWilsonCoefficient}. The fact that the sixth order still gives rise to a unique non-trivial fixed point with just one relevant direction points towards an Asymptotic Safety scenario of pure quantum gravity that features no free parameters. Compared to the standard scheme which usually gives rise to three to four relevant directions \cite{Codello:2008vh,Benedetti:2009gn,Falls:2014tra,Denz:2016qks,Falls:2017lst,Kluth:2020bdv, Falls:2020qhj,Sen:2021ffc}, we see that identifying redundant couplings is key to understand the predictivity of the fixed point within Asymptotic Safety.

We have also observed several features of near-perturbativity at the non-trivial fixed point. The clearest indication is the very small value of the Goroff-Sagnotti coupling at the fixed point, $g_{C^3} = 4.5 \cdot 10^{-7}$, see \eqref{eqn:fpcoup}. Additionally, we have shown in \eqref{eqn:gammas} that the \ac{RG} kernel which parameterises the field redefinitions takes small absolute values as well, with the sole exception of $\gamma_g$ taking values of the order of one. Lastly, critical exponents receive minor corrections when compared to their canonical counterparts. All these indications line up with previous observations \cite{Falls:2013bv,Falls:2014tra, Eichhorn:2018akn, Eichhorn:2018ydy}, and suggest a near-perturbative behaviour of quantum gravity. It is as non-perturbative as necessary, but as perturbative as it gets.

This observation can lead to possible simplifications for future investigations: a possible way to compute higher order corrections is to keep all essential couplings in the action to a given order, while only retaining $\gamma_g$ in the \ac{RG} kernel. Remarkably, implementing this seemingly crude approximation, the results at sixth order in the derivative expansion are virtually unchanged, see \eqref{eqn:simple}. This is to be contrasted with the standard scheme, where such an approximation leads to a strongly non-canonical irrelevant critical exponent, see \cite{Gies:2016con} and appendix \ref{app:C3standard}.

Finally, we have checked the dependence of our results on technical parameters. This serves as an important tool to estimate truncation errors that are present in applications of the \ac{FRG}. Our analysis shows that our results are largely independent of such technical choices. We have tested this for two different shape functions and with two different technical parameters: $\alpha$, which enters as a prefactor in front of the shape function of the regulator \eqref{eqn:regdef}, and $\beta$, which parameterises the endomorphism of the Laplacian \eqref{eqn:endodef}. A sophisticated numerical analysis shows that qualitative properties are independent of technical parameters. As seen in \autoref{fig:PMSRel}, a large area is found in which critical exponents plateau within which we find a point of minimum sensitivity with respect to both parameters, see \eqref{eqn:PMSLitim1} and \eqref{eqn:PMSLitim2}. Outside of this area, stronger fluctuations are present. We expect that in this regime, our approximation scheme breaks down. Most importantly, the critical exponent of the Goroff-Sagnotti coupling is always irrelevant, and the relevant critical exponent takes mostly values of $\theta_1 \approx 2.1 \pm 0.2$. These results also hold for different shape functions. In this work, we compared the Litim cutoff and its ``square'', and find qualitative agreement. This analysis also showed the special role played by the type II cutoff $(\beta = 1)$. For this choice, we find the strongest indications of near-perturbativity. Except for $\gamma_g$, all $\gamma$-functions that parameterise the \ac{RG} kernel develop a zero close to $\beta = 1$, see \autoref{fig:gammafuncs}. \autoref{fig:gC3} shows that this also applies to the fixed point value of the Goroff-Sagnotti coupling which develops a zero close to $\beta = 1$ as well.

It will be of great interest to apply our findings to future endeavours, including even higher orders in the derivative expansion. This might be particularly feasible due to the near-perturbativity of the fixed point which potentially allows for much simplified yet accurate approximations. Similarly, the application of this scheme to gravity-matter systems is intriguing. First steps have been made in \cite{Knorr:2022ilz}. Observing the fixed point to persist the inclusion of matter would be a strong indication for a unitary and \ac{UV}-complete theory of quantum gravity.
By studying gravity-matter systems it would also be interesting to compute the scaling dimension of relational observables \cite{Baldazzi:2021fye} at the higher orders of the derivative expansion.
Last but not least, a critical open question is the extension of the \ac{MES} to setups beyond the background field approximation. This can be explored within the fluctuation approach \cite{Pawlowski:2023gym}, within perturbation theory \cite{Martini:2022sll} or within a manifestly background independent approach such as \cite{Falls:2020tmj}.

\begin{acknowledgments}

The authors would like to thank Roberto Percacci for useful comments on the manuscript. YK was supported by a United Kingdom Research and Innovation (UKRI) Future Leaders Fellowship [Grant No.~MR/V021974/2]. YK thanks SISSA for hospitality during early stages of this project. YK is grateful for the hospitality of Perimeter Institute where part of this work was carried out. Research at Perimeter Institute is supported in part by the Government of Canada through the Department of Innovation, Science and Economic Development and by the Province of Ontario through the Ministry of Colleges and Universities. Nordita is supported in part by NordForsk. For the purpose of open access, the authors have applied a Creative Commons Attribution (CC BY) licence to any Author Accepted Manuscript version arising.

\end{acknowledgments}

\section*{Data Access Statement}
We provide a Mathematica notebook with all $\beta$- and $\gamma$-functions for arbitrary shape functions and technical parameters \cite{SupplementalMaterial}. No other data were created or analysed in this study.

\appendix

\section{The Two-Loop counterterm in the standard scheme}\label{app:C3standard}

With our setup, it is straightforward to try to reproduce the original \ac{FRG} computation that included just the Einstein-Hilbert action amended by the Goroff-Sagnotti counterterm \cite{Gies:2016con}. We found some inaccuracies in the $\beta$-function for $g_{C^3}$ (called $\sigma$ in the reference).\footnote{There is also a typo in the expression for $B_1$ (the last term in the brackets should be $-7$ instead of $-5$), but the numerics were done with the correct expression for $B_1$.} Borrowing a similar notation, we parameterise the $\beta$-function as
\begin{equation}
    \partial_t g_{C^3} = c_0 + (2 + c_1) g_{C^3} + c_2 g_{C^3}^2 + c_3 g_{C^3}^3 \, .
\end{equation}
With the standard notation $\eta_N=\frac{\partial_t g - 2g}{g}$, the correct coefficients $c_i$ read
\begin{equation}
    \begin{aligned}
        c_0 &= \frac{1}{64\pi^2(1-2\lambda)} \left( \frac{2-\eta_N}{2(1-2\lambda)} + \frac{3+\eta_N}{(1-2\lambda)^3} - \frac{\eta_N}{756} \right) \, , \\
        c_1 &= \frac{48g}{16\pi(1-2\lambda)^2} \bigg( 45(6-\eta_N) + 59 \frac{8-\eta_N}{1-2\lambda} \\
        &\hspace{5cm} + \frac{63}{10} \frac{10+\eta_N}{(1-2\lambda)^2} \bigg) \, , \\
        c_2 &= \frac{g^2}{2(1-2\lambda)^3} \left( \frac{43}{5} (12-\eta_N) - \frac{9}{7} \frac{112-\eta_N}{1-2\lambda} \right) \, , \\
        c_3 &= 6\pi g^3 \frac{18-\eta_N}{(1-2\lambda)^4} \, .
    \end{aligned}
\end{equation}
Despite these corrections, the overall picture is extremely similar to the one reported in \cite{Gies:2016con}. On the qualitative level, this is obvious due to the general arguments given in \cite{Gies:2016con} about the cubic structure of the above $\beta$-function. Quantitatively, we again only find two real fixed points. The shifted Gaussian fixed point is at
\begin{equation}
    g_N^\ast=\lambda^\ast=0 \, , \qquad g_{C^3}^\ast = \frac{1}{64\pi^2} \, ,
\end{equation}
with canonical critical exponents
\begin{equation}
    \theta^{sGFP} \in \left\{ 2,-2,-2 \right\} \, .
\end{equation}
The non-trivial, interacting fixed point now sits at
\begin{equation}
    g_N^\ast=0.707 \, , \quad \lambda^\ast=0.193 \, , \quad g_{C^3}^\ast = 1.368 \times 10^{-4} \, ,
\end{equation}
with critical exponents
\begin{equation}
    \theta_{1,2} = 1.475 \pm 3.043 \, \mathbf{i} \, , \qquad \theta_3 = -20.097 \, .
\end{equation}
The irrelevant critical exponent is still rather large compared to the one that we find in the \ac{MES}, independent of whether we do or do not take into account all operators up to six powers in derivatives.

\begin{figure*}
    \centering
    \begin{minipage}[t]{0.49\textwidth}
        \large Fourth Order \\[1ex]
        \includegraphics[width=\textwidth]{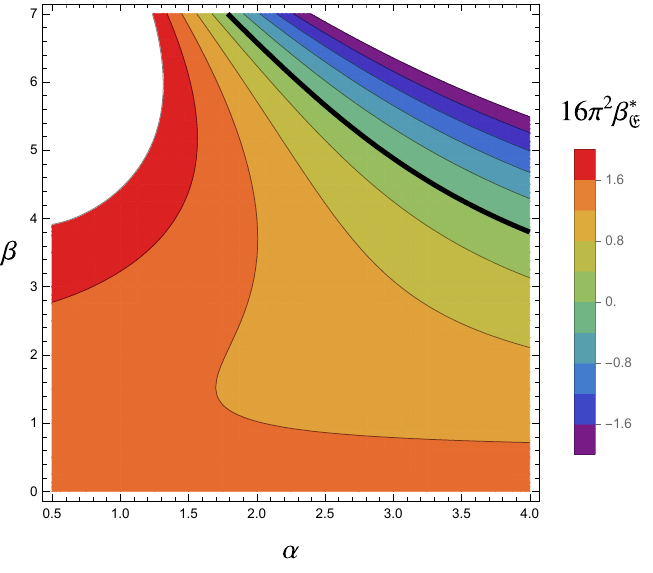}
    \end{minipage}
    \hfill
    \begin{minipage}[t]{0.49\textwidth}
        \large Sixth Order \\[1ex]
        \includegraphics[width=\textwidth]{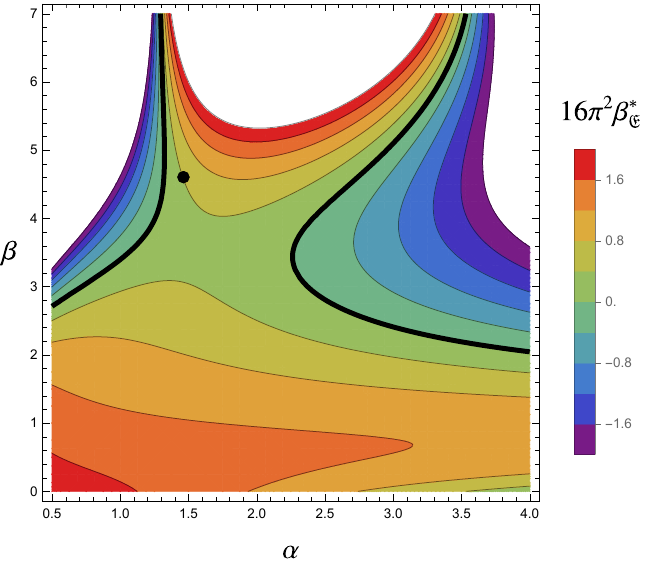}
    \end{minipage}
    \caption{\label{fig:betaeuler}The value for the $\beta$-function of the topological invariant at the Reuter fixed point in dependence of $\alpha$ and $\beta$ with the Litim cutoff. The left panel shows the fourth order and the right panel the sixth order in derivatives. White areas denote either instabilities of our approximation, or areas where $\beta_\eulerterm$ takes values outside the plotted domain.}
\end{figure*}

\section{The Topological Term}\label{app:GB}

We have already mentioned in the main text that the fourth order of the derivative expansion features the topological Gauss-Bonnet term,
\begin{equation}
    \eulerterm = R_{\rho \sigma \mu \nu} R^{\rho \sigma \mu \nu} - 4 R_{\mu \nu} R^{\mu \nu} + R^2 \, ,
\end{equation}
whose spacetime integral is related to the Euler characteristic,
\begin{equation}
    \chi = \frac{1}{32 \pi^2} \int \dd^d x \sqrt{g} \, \eulerterm \, .
\end{equation} 
Due to its topological nature, \eulerterm~is locally a total derivative. Hence, the Gauss-Bonnet coupling does not appear in any $\beta$-functions of the theory. A consequence of this is that the $\beta$-function of the Gauss-Bonnet ($\beta_\eulerterm$) is independent of itself and receives only contributions from other couplings in the theory. Even though these other couplings can have a non-trivial fixed point, the Gauss-Bonnet coupling only stops running if its $\beta$-function vanishes due to non-trivial cancellations between the contributions of other couplings. However, we are not aware of any arguments why such a cancellation would take place.

If $\beta_\eulerterm$ does not vanish at the non-trivial fixed point, the Gauss-Bonnet coupling either runs to $+ \infty$ or $- \infty$. In \cite{Knorr:2021slg}, it was argued that this leads to a suppression/enhancement of certain topologies in the path integral. Depending on the sign of $\beta_\eulerterm$, the contribution of topologies with a given sign of the Euler characteristic is enhanced over the other. 

We present our findings for $\beta_\eulerterm$ at the non-trivial fixed point in \autoref{fig:betaeuler} in dependence on $\alpha$ and $\beta$. To understand the convergence of the result for $\beta_\eulerterm^\ast$, we show the results at fourth and sixth order in the derivative expansion. Comparing both orders, we conclude that the biggest stability for $\beta_\eulerterm^\ast$ is found for the type II cutoff, \ie{} $\beta = 1$. This is in line with our findings in the main text, see \eg{} \autoref{fig:gammafuncs}. In this regime, we find a positive value for $\beta_\eulerterm^\ast$. For $\alpha = \beta = 1$ we have 
\begin{equation}\label{eqn:betaEuvfp}
    16 \pi^2 \beta_\eulerterm^\ast \approx 1.273 \, .
\end{equation}
This indicates that the Gauss-Bonnet coupling runs towards $+\infty$ in the \ac{UV}. In the Euclidean path integral, this means that the weight of topological configurations with positive Euler characteristic is suppressed, while topologies with negative Euler characteristics are enhanced. We also find parameter regions with vanishing or negative $\beta_\eulerterm^\ast$, however, comparing fourth and sixth orders, we find that our approximation is less stable in these ranges. Thus, our results suggest a positive $\beta_\eulerterm$ at the non-trivial fixed point, in line with previous findings \cite{Knorr:2021slg, Knorr:2022ilz, Knorr:2023usb}.

These findings can also be compared to the value of $\beta_\eulerterm$ at the Gaussian fixed point. There, we rediscover the known result
\begin{equation}
    16 \pi^2 \beta_\eulerterm^\text{Gauss} = \frac{53}{45} \approx 1.178\, .
\end{equation}
Within the uncertainties of our computation, this is identical to \eqref{eqn:betaEuvfp} at the non-trivial fixed point. The topological $\beta$-function is virtually unchanged at the non-trivial fixed point, similar to the results at fourth order in derivatives \cite{Baldazzi:2021orb}.

\bibliography{ess_cubic}

\end{document}